\documentclass[english]{article}
\usepackage[T1]{fontenc}
\usepackage[latin9]{inputenc}
\usepackage{babel}
\usepackage{graphicx,amsmath,amssymb,color}
\usepackage[normalem]{ulem}
\usepackage{amsfonts}
\usepackage[toc,page]{appendix}
\usepackage{hyperref}
\usepackage{latexsym}
\usepackage{amsfonts}
\usepackage{algpseudocode}
\usepackage{amsthm}
\usepackage{mathrsfs}
\usepackage{color,verbatim}
\usepackage{psfrag}
\usepackage{rotating} 
\usepackage{bbold} 
\usepackage{multirow}

\usepackage{subcaption} 

\usepackage[svgnames]{xcolor}
\usepackage{tikz}

\usepackage{algorithm}

\newcommand{\bra}[1]{\langle#1 |}
\newcommand{\ket}[1]{|#1 \rangle}
\newcommand{\bigket}[1]{\Bigl \lvert#1  \Bigr \rangle}
\newcommand{\braket}[2]{\left \langle #1 \middle \vert #2 \right \rangle}
\newcommand{\bigbraket}[2]{\left \langle #1 \middle\vert #2 \right \rangle}
\newcommand{\ketbra}[2]{\vert #1 \rangle \! \langle #2 \vert}
\newcommand{\overlap}[2]{\left \langle #1 | #2 \right\rangle}
\newcommand{\average}[1]{\langle #1 \rangle}
\newcommand{\sandwich}[3]{\left \langle #1 \middle \vert #2 \middle \vert #3 \right\rangle}
\newcommand{\innerprod}[2]{\left \langle #1 | #2 \right\rangle}
\newcommand{\nick}[1]{\textcolor{blue}{Nick: #1}}
\definecolor{myDarkRed}{rgb}{0.5, 0, 0} 
\newcommand{\catherine}[1]{\textcolor{myDarkRed}{Catherine: #1}}
\newcommand{\uchenna}[1]{\textcolor{magenta}{Uchenna: #1}}
\definecolor{myOrange}{rgb}{0.7, 0.3, 0} 
\definecolor{myDarkGreen}{rgb}{0, 0.7, 0} 
\newcommand{\raouf}[1]{\textcolor{myDarkGreen}{Raouf: #1}}
\definecolor{myAqua}{rgb}{0, 0.3, 0.3}
\newcommand{\jesse}[1]{\textcolor{myOrange}{Jesse: #1}}
\newcommand{\removed}[1]{\textcolor{red}{\sout{#1}}}
\newcommand{\added}[1]{\textcolor{blue}{#1}}

\title{Zeno-effect Computation: Opportunities and Challenges}
\author{Jesse Berwald$^\dagger$, Nicholas Chancellor, Raouf Dridi$^\dagger$}
\date{Quantum Computing Inc (QCi), 5 Marine View Plaza Hoboken NJ 07030 USA \newline
$\dagger$ current affiliation: Qamia Quantum Technologies
Albuquerque, USA and Abu Dhabi, UAE \\ \today}

\def\justbeingincluded{justbeingincluded} 

\begin{document}

\ifx \justbeingincluded \undefined

\documentclass[english]{article}
\usepackage[T1]{fontenc}
\usepackage[latin9]{inputenc}
\usepackage{babel}
\usepackage{graphicx,amsmath,amssymb,color}
\usepackage[normalem]{ulem}
\usepackage{amsfonts}
\usepackage[toc,page]{appendix}
\usepackage{hyperref}
\usepackage{latexsym}
\usepackage{amsfonts}
\usepackage{algpseudocode}
\usepackage{amsthm}
\usepackage{mathrsfs}
\usepackage{color,verbatim}
\usepackage{psfrag}

\usepackage{rotating} 
\usepackage{bbold} 
\usepackage{multirow}

\usepackage{subcaption} 

\usepackage[svgnames]{xcolor}
\usepackage{tikz}

\usepackage{algorithm}

\newcommand{\bra}[1]{\langle#1 |}
\newcommand{\ket}[1]{|#1 \rangle}
\newcommand{\bigket}[1]{\Bigl \lvert#1  \Bigr \rangle}
\newcommand{\braket}[2]{\left \langle #1 \middle \vert #2 \right \rangle}
\newcommand{\bigbraket}[2]{\left \langle #1 \middle\vert #2 \right \rangle}
\newcommand{\ketbra}[2]{\vert #1 \rangle \! \langle #2 \vert}
\newcommand{\overlap}[2]{\left \langle #1 | #2 \right\rangle}
\newcommand{\average}[1]{\langle #1 \rangle}
\newcommand{\sandwich}[3]{\left \langle #1 \middle \vert #2 \middle \vert #3 \right\rangle}
\newcommand{\innerprod}[2]{\left \langle #1 | #2 \right\rangle}
\newcommand{\nick}[1]{\textcolor{blue}{Nick: #1}}
\definecolor{myDarkRed}{rgb}{0.5, 0, 0} 
\newcommand{\catherine}[1]{\textcolor{myDarkRed}{Catherine: #1}}
\newcommand{\uchenna}[1]{\textcolor{magenta}{Uchenna: #1}}
\definecolor{myOrange}{rgb}{0.7, 0.3, 0} 
\definecolor{myDarkGreen}{rgb}{0, 0.7, 0} 
\newcommand{\raouf}[1]{\textcolor{myDarkGreen}{Raouf: #1}}
\definecolor{myAqua}{rgb}{0, 0.3, 0.3}
\newcommand{\jesse}[1]{\textcolor{myOrange}{Jesse: #1}}
\newcommand{\removed}[1]{\textcolor{red}{\sout{#1}}}
\newcommand{\added}[1]{\textcolor{blue}{#1}}

\title{Zeno-effect Computation: Opportunities and Challenges}
\author{Jesse Berwald$^\dagger$, Nicholas Chancellor, Raouf Dridi$^\dagger$}
\date{Quantum Computing Inc (QCi), 5 Marine View Plaza Hoboken NJ 07030 USA \newline
$\dagger$ current affiliation: Qamia Quantum Technologies
Albuquerque, USA and Abu Dhabi, UAE \\ \today}

\begin{document}

\fi 

\maketitle

\abstract{
Adiabatic quantum computing has demonstrated how quantum Zeno can be used to construct quantum optimisers. However, much less work has been done to understand how more general Zeno effects could be used in a similar setting. We use a construction based on three state systems rather than directly in qubits, so that a qubit can remain after projecting out one of the states. We find that our model of computing is able to recover the dynamics of a transverse field Ising model, several generalisations are possible, but our methods allow for constraints to be implemented non-perturbatively and does not need tunable couplers, unlike simple transverse field implementations. We further discuss how to implement the protocol physically using methods building on STIRAP protocols for state transfer. We find a substantial challenge, that settings defined exclusively by measurement or dissipative Zeno effects do not allow for frustration, and in these settings pathological spectral features arise leading to unfavorable runtime scaling. We discuss methods to overcome this challenge for example including gain as well as loss as is often done in an optical setting. 
}

\tableofcontents

\section{Introduction}

The use of Zeno effects in quantum computation has traditionally focused on either adiabatic optimisation \cite{farhi00a,albash16a} or on implementation within a gate model setting, (for example universal Zeno-based gate sets) \cite{ Beige2000dissdecohere,Huang2008zenouniversal,Xu2009ZenoPhase,Huang2012Fredkin,Sun2013ZenoGates,dizaji2024zenosubspaces,liu2024comparisonconstrainencodingmethods}. These are both promising areas, and in particular quantum annealing, a generalisation of adiabatic quantum annealers beyond the adiabatic regime where the computation can be described as a simple Zeno effect has proven to be a highly promising field \cite{kadowaki98a,Crosson14a,crosson2020prospects,Callison21a,Yarkoni2022AnnealingIndustry,Au-Young2023QuantumOpt,Banks2024continuoustime,Banks2024rapidquantum,Schulz2023GuidedQW}. 

While open quantum system effects can play a useful role in annealing computation \cite{dickson13a}, these effects are usually manifested as thermal dissipation rather than Zeno effects. Dissipative open system effects are in fact even the driving mechanism behind the form of reverse annealing which can be experimentally implemented in flux qubit devices \cite{chancellor17b,Callison2022hybrid}. 

In optical computation, coherent and spatial photonic Ising machines can be implemented by balancing gain and loss such that the most optimal solution is amplified at the expense of the others \cite{McMahon16a,Inagaki16a,Yamamoto2017CoherentIsing,Kumar2020opticalIsing,Pierangeli2020SPIM,Yamamoto2020CoherentIsing,Prabhakar2023photonic}. These devices again operate on different mechanisms than Zeno effects, instead being based on the tendency of non-linear optical systems to preferentially populate modes (or quadratures) which are already the most populated, the same physical mechanism which leads to the well known Hong-Ou-Mandel effect \cite{Hong1987HOM}. 

An area which is much less explored, but very promising is how more general Zeno effects, beyond the well explored adiabatic setting could be used to compute. This area does have promise, It was recently shown \cite{Berwald2024Zeno}, that analog optimisers based on general Zeno effects\footnote{This had previously been shown in the adiabatic setting by Roland and Cerf \cite{Roland2002}} can yield an equivalent speedup to Grover's well known gate-model algorithm for unstructured search \cite{Grover1996search,Grover1997Search}. Grover's algorithm represents the best possible performance for unstructured search on a quantum device \cite{Bennett1997GroverBest}. A key motivation of that work was to provide theoretical grounding for emerging paradigms, base on novel Zeno effects, in particular entropy computing \cite{nguyen2024entropycomputing}. However, \cite{Berwald2024Zeno} did not examine how to solve practical problems, as we do here. Similarly, general error bounds have been derived for different types of Zeno effects \cite{Burgarth2022oneboundtorulethem}.

Within this work we explore how to compute using Zeno effects beyond the traditional and well explored adiabatic setting. We particularly are interested in how dissipative and measurement Zeno effects can be used to solve optimisation problems. We focus on the case where the system remains in a pure state, and the feedback which is usually used in measurement based quantum computation is not present \cite{Briegel2009MBQC}. An important challenge which is not present when Zeno effects are implemented adiabatically is that these implementations do not directly support frustration.

A frustrated system is defined as one where the lowest energy state of a sum of Hamiltonian terms does not minimize each of the terms individually. A simple example of frustration which is often used is a triangle of anti-ferromagnetic Ising couplings.

\begin{equation}
H_3=Z_1Z_2+Z_2Z_3+Z_1Z_3
\end{equation}

In such a system each coupling individually is satisfied if the two connecting spins disagree, but there is no way to assign values to all three spins such that each of the three pairwise combinations disagree. The ground state of such a system would be a six-fold degenerate combination of $\ket{001}$, $\ket{010}$, $\ket{100}$, $\ket{110}$, $\ket{101}$ and $\ket{110}$, excluding only the two states where all three agree.

Trying to develop an analog to frustration but in a destructive setting (in the sense of \cite{Berwald2024Zeno} where the system is sent to a computationally useless state $\ket{d}$ under certain conditions) leads to problems. Taking the triangle example, a dissipative analog would be to couple all states with any $\ket{00}$ and $\ket{11}$ pairs to a destroyed state $\ket{d}$. All possible three qubit states will contain at least one $\ket{00}$ or $\ket{11}$ pair. Therefore in the limit of strong coupling which is needed for Zeno effects to be implemented, such a system would always end in a computationally useless state. 

Similarly, if one tried to implement frustration with a measurement based Zeno effect, measurements in incompatible bases would lead to unavoidable decoherence. In the case of the triangle example, this would mean measuring this would mean measuring $\average{Z_1Z_2}$, $\average{Z_2Z_3}$, and $\average{Z_1Z_3}$. Such measurements would lead to loss of coherence between states where different pairs disagree, for example between $\ket{110}$ and $\ket{101}$. In $\ket{110}$ we would find $\average{Z_1Z_2}=1$ while $\average{Z_2Z_3}=\average{Z_1Z_3}=-1$, while in $\ket{101}$ we would find $\average{Z_1Z_3}=1$ while $\average{Z_1Z_2}=\average{Z_2Z_3}=-1$. Different classical measurement patterns mean that these states would be distinguishable and therefore the state $(\ket{110}+\ket{101})/\sqrt{2}$ would be subject to decoherence.

The reason adiabatic Zeno effects support frustration, while others do not is that quantum mechanics is invariant under global phase rotations. Being in a higher energy state of one (or more) Hamiltonian term only contributes a global phase to the ground state. In contrast replacing this term by a measurement or destruction operation, is highly disruptive to the underlying system.

To explore what is possible when solving optimisation problems in a more general Zeno setting, we construct a simple mathematical implementation of Zeno effect computation, where the fundamental subunit is a three-level system in which a time-dependent state is projected out by a Zeno effect. We then combine this system with a problem statement taking the form of either time evolution with a Hamiltonian, or constraints which are implemented through a different Zeno effect. We show that if perfect Zeno projection is performed in this three-state generalisations of Pauli $Z$ operators, the physics of a transverse-field Ising model are recovered. We also discuss several generalisations.

Within this paper we consider examples where the Zeno-projection subunits are combined with constraints, for example requiring that clauses of a three satisfiability problem are not violated. We find that our formalism does indeed lead to a valid subspace which does not couple to dissipation if and only if all constraints are simultaneously satisfiable. We however find that in practice when implemented dissipatively, this leads to a number of eigenstates which couple only very weakly to dissipation  and therefore will display only weak Zeno effects. We show numerically that a consequence of this feature is that the runtime is very long to approach a success probability of order one. While such a feature does not completely rule out the usefulness of these protocols it is probably undesirable. We further show that the same success probability feature arises when implemented using measurements. This is likely because both protocols become mathematically identical in the limit of many measurements or strong dissipative coupling. 

There are at least three possible methods to avoid the pathological behaviour we see here, one of which we explore numerically in this work, and two others which have been discussed elsewhere:
\begin{enumerate}
    \item Add some frustration through additional Hamiltonian terms. We demonstrate this using what we call an ``offset'' Hamiltonian. An example of which is given here.
    \item Consider systems with gain as well as loss, as is often done in an optical setting \cite{Kumar2020opticalIsing,Yamamoto2017CoherentIsing,Yamamoto2020CoherentIsing,Honjo2021CoherentIsing,Mosheni2022IsingMachines,Lu2023CoherentIsing,nguyen2024entropycomputing}. This requires many-body formalism which is beyond the scope of this work.
    \item Directly implement measurement or dissipation in a complex frustrated basis. This requires a relatively complex coupling between the computational units and another system, such implementations were explored in \cite{Berwald2024Zeno}; but will not be explored here.
\end{enumerate}

We further highlight that systems engineered using our formulation implement constraints in a fundamentally non-perturbative way. This has an advantage over traditional implementations based on adding very strong couplings to a transverse field Ising model. In that case strengthening the constraints effectively slows the dynamics leading to a tradeoff between performance in finding the lowest energy state of the Ising model and faithfully defining the problem to be solved. We argue analytically and demonstrate numerically that a tradeoff where stronger constraints lead to slower dynamics does not exist when constraints are implemented within the three-state formalism we develop here. We further show that an experimental implementation of the system here could be constructed by building upon the well known STIRAP, (stimulated Raman adiabatic passage, \cite{Vitnov2017STIRAP}) protocol which is often used in atomic physics.

\subsection{Paper structure}

Since we report a number of substantially different but related results, it is worth discussing the overall structure of our work and the key results of each section. Firstly, section \ref{sec:formalism} gives the underlying formalism which we use for the remainder of the work. This sets the scene and provides the necessary definitions to understand the later results. We argue that this is likely to be the simplest way to define qubit computation which can be achieved by general Zeno effects. 

Next in section \ref{sec:eff_Ising} we perform a first investigation of the mathematical structure. For this section, we assume that whatever Zeno effect is used, it is able to perfectly project into the manifold of allowed states. Our main finding in this section is that under these assumptions, a transverse field Ising model can be recovered by applying an energy offset to a state we call the ``undefined'' state of each computational unit.

In section \ref{sec:zeno_inc}, we examine how well specific implementations of Zeno effects perform. We find that a mathematical structure arises where the Zeno effects will be very weak at the beginning of the protocol. We further see that that this structure manifests itself as poor scaling of success probability with runtime. We find that this structure is removed if the same kind of energy offset which allowed a transverse Ising model to be realized in section \ref{sec:eff_Ising} is added. We argue that this is evidence of the importance of frustration, as this term could not be accomplished by using pure Zeno projection. When implemented as a Hamiltonian this means the ground state cannot satisfy all individual Hamiltonian terms simultaneously.

Next in section \ref{sec:adiabat_num} we show a specific example of how, in an adiabatic implementation of the formalism discussed here constraints can be implemented in a way which allows dynamics when constraints are made arbitrarily strong. This is in contrast to the traditional implementation of constraints in annealing systems, which is ``perturbative'' in the sense that the system must pass through at least one very high energy configuration to explore the valid solutions. For this reason we refer to these as ``non-perturbative'' constraints.

In section \ref{sec:exp_imp} we discuss how the systems discussed here could be constructed experimentally. This construction has a strong connection to an atomic physics protocol known as STIRAP. We also discuss at a high level the implications that such an implementation would have in terms of errors. This section is followed by a brief section on numerical methods for reproducibility purposes \ref{sec:num_meth}. Finally, there are discussions and conclusions of the work \ref{sec:discuss_conclude}. 

In summary, we see the four key results (in the order they are discussed) as:

\begin{itemize}
    \item Capability of our formalism to reproduce a transverse Ising model as reported in section \ref{sec:eff_Ising}
    \item Mathematical structures which inhibit performance when frustration is absent as reported in section \ref{sec:zeno_inc}
    \item Ability to implement non-perturbative constraints within three-state systems as reported in section \ref{sec:adiabat_num}
    \item Experimental implementation of the abstract system proposed here as reported in section \ref{sec:exp_imp}
\end{itemize}

sections not mentioned in these bullet points provide supporting material for these key results as well as containing some secondary results.

\section{Formulation \label{sec:formalism}}

Qubits form a simple unit for performing computations, but to effectively make use of Zeno physics the Hilbert space needs to be expanded to include at least one state which can then be forbidden using a Zeno effect, leaving a non-trivial two-state subspace as opposed to a trivial one-state one. For this reason we consider tensor product structures built from units with 
three levels. One such structure consists of the space spanned by $\{\ket{0},\ket{1},\ket{u}\}$. The $\ket{0}$ and $\ket{1}$ states are the usual qubit states, while $\ket{u}$ is a third ``undefined'' state. Conceptually the computational protocol we consider consists of initialising in the undefined state $\ket{u}$ and continuously evolving the space until the variable is forced to be in the space spanned by $\{\ket{0},\ket{1}\}$.

Mathematically, the desired behaviour can be achieved by coupling to 
\begin{equation}
    \ket{\xi(\theta)}=-\sin(\theta)\ket{u}+\cos(\theta)\ket{+}, \label{eq:coupling_state}
\end{equation}
and continuously evolving $\theta=0\rightarrow\theta=\pi/2$. Note that we use the conventional notation $\ket{\pm}=(\ket{0}
\pm\ket{1})/\sqrt{2}$.  Assuming strong enough coupling to achieve a Zeno effect, the system is confined to the orthogonal complement of $\ket{\xi(\theta)}$, a two-dimensional space spanned by 
\begin{equation}
    \{\ket{\tilde{+}}=\cos(\theta)\ket{u}+\sin(\theta)\ket{+},\ket{\tilde{-}}=\ket{-}\},\label{eq:allowed_span}
\end{equation}
or equivalently
\begin{align}
 \{\ket{\tilde{0}}=\frac{1}{\sqrt{2}}\left(\cos(\theta)\ket{u}+\sin(\theta)\ket{+}+\ket{-}\right),\nonumber \\
 \ket{\tilde{1}}=\frac{1}{\sqrt{2}}\left(\cos(\theta)\ket{u}+\sin(\theta)\ket{+}-\ket{-}\right)\}   \label{eq:zero_one_tilde}
\end{align}
for all values of $\theta$. Since $\braket{\xi(\theta)}{-}=0\, \forall \theta$ and $\braket{\xi(\theta)}{\xi(\theta+\epsilon)}\approx 1\, \forall \theta$, this system will perform a continuous rotation between the $\ket{u}$ and $\ket{+}$ state. If we tensor many such systems together, then a sweep in $\theta$ will result in a uniform positive superposition of all possible solutions.

\begin{figure}
    \centering
    \includegraphics[width=10 cm]{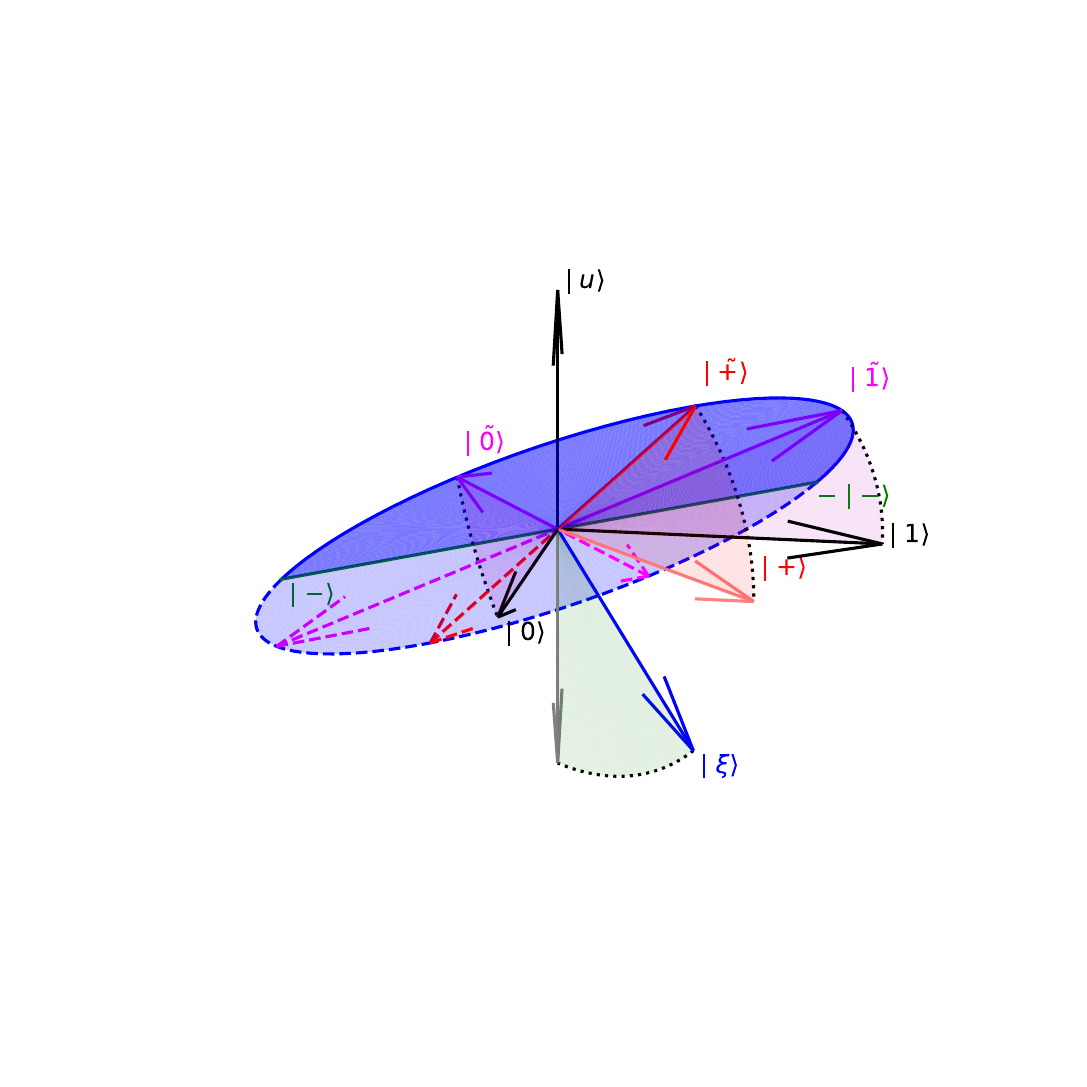}
    \caption{Visualisation of the relationship between $\ket{\xi}$,$\ket{\tilde{0}}$,$\ket{\tilde{1}}$, and $\ket{\tilde{+}}$ and the states $\ket{0}$, $\ket{1}$ and $\ket{u}$ which form the basis for $\theta=\pi/4$. The blue disk indicates allowed superpositions, with the back half of the disk showing redundant states which only differ from the front half by a global phase and are therefore quantum mechanically indistinguishable. The grey vector pointing downward is a guide to the eye, as are dashed lines and shading of angles. Dashed vectors indicate versions which are transformed by a global minus sign. The orthogonality relationships between states is summarised in table \ref{tab:orthogonality_relationships}.}
    \label{fig:3d_visualisation}
\end{figure}

If we neglect the possibility of complex superpositions, in other words those which involve relative factors of $i=\sqrt{-1}$, then we can visualise the rotation we perform here in the three-dimensional space spanned by $\{ \ket{u},\ket{0},\ket{1}\}$. Since all vectors we have defined here are purely real such visualisations will not show every possible superposition. Nevertheless, they will prove valuable to visualise the states relevant to the computatation we perform here. Figure \ref{fig:3d_visualisation} illustrates this vector space, with orthogonality relationships summarised in table \ref{tab:orthogonality_relationships}. From this visualisation several facts become clear. Firstly, while $\ket{\tilde{0}}$ and $\ket{\tilde{1}}$ are orthogonal to each other it is not generally true that $\ket{\tilde{0}}$ and $\ket{1}$ will be orthogonal or vice versa. Furthermore, given the nature of the rotation, $\ket{-}$ is the only state which lies within the allowed space for all $\theta$.

To perform meaningful computation, additional interactions need to be introduced between the $\ket{+}$ and $\ket{-}$ states. This can either be accomplished by acting with $Z=\ketbra{0}{0}-\ketbra{1}{1}$, or by including dissipative coupling to induce further Zeno effects. We show that the former can be used to construct an effective transverse Ising system, while the latter can be used to construct a solver for finding qubit arrangements which satisfy constraints.

\begin{table}[]
    \centering
    \begin{tabular}{|c||c|c|c|c|c|c|c|c|c|}
    \hline
         & & & & & & & & & \\[-10 pt] 
        state & $\ket{0}$ & $\ket{1}$ & $\ket{\tilde{+}}$ & $\ket{\tilde{-}}$ & $\ket{\tilde{0}}$ & $\ket{\tilde{1}}$ & $\ket{\xi}$ & $\ket{u}$ & $\ket{\phi} \,(\perp \ket{0})$
        \\
        \hline
        \hline
        & & & & & & & & & \\[-10 pt]
        $\ket{0}$ & - & $\forall \theta$ & $0$ & - &  - & $\frac{\pi}{2}$ & $\frac{\pi}{2}$ & $\forall \theta$ & $\forall \theta$ \\
        & & & & & & & & & \\[-10 pt] 
        \hline
        & & & & & & & & & \\[-10 pt]
        $\ket{1}$ & $\forall \theta$ & - & $0$ & - &   $\frac{\pi}{2}$ & - & $\frac{\pi}{2}$ & $\forall \theta$ & $0$ \\
        & & & & & & & & & \\[-10 pt]
        \hline
        & & & & & & & & & \\[-10 pt]
        $\ket{\tilde{+}}$ & $0$ & $0$ & - & $\forall \theta$ & - & - & $\forall \theta$ & $\frac{\pi}{2}$ & -
        \\
        & & & & & & & & & \\[-10 pt]
        \hline
        & & & & & & & & & \\[-10 pt]
        $\ket{\tilde{-}}$ & - & - & $\forall \theta$ & - & - & - & $\forall \theta$ & $\forall \theta$ & $0$
        \\
        & & & & & & & & & \\[-10 pt]
        \hline
        & & & & & & & & & \\[-10 pt]
        $\ket{\tilde{0}}$ & - & $\frac{\pi}{2}$ & - & - & - & $\forall \theta$ & $\forall \theta$ & $\frac{\pi}{2}$ & $\frac{\pi}{2}$
        \\
        & & & & & & & & & \\[-10 pt]
        \hline
        & & & & & & & & & \\[-10 pt]
        $\ket{\tilde{1}}$ & $\frac{\pi}{2}$ & - & - & - & $\forall \theta$ & - & $\forall \theta$ & $\frac{\pi}{2}$ & 0
        \\
        & & & & & & & & & \\[-10 pt]
        \hline
        & & & & & & & & & \\[-10 pt]
        $\ket{\xi}$ & $\frac{\pi}{2}$ & $\frac{\pi}{2}$ & $\forall \theta$ & $\forall \theta$ & $\forall \theta$ & $\forall \theta$ & - & $0$ & $\forall \theta$ \\
        & & & & & & & & & \\[-10 pt]
        \hline
        & & & & & & & & & \\[-10 pt]
        $\ket{u}$ & $\forall \theta$ & $\forall \theta$ & $\frac{\pi}{2}$ & $\forall \theta$ & $\frac{\pi}{2}$ & $\frac{\pi}{2}$ & $0$ & - & $\frac{\pi}{2}$
        \\
        & & & & & & & & & \\[-10 pt]
        \hline
        & & & & & & & & & \\[-10 pt]
        $\ket{\phi} \,(\perp \ket{0})$ & $\forall \theta$ & $0$ & - & $0$ & $\frac{\pi}{2}$ & -  & $\forall \theta$ & $\frac{\pi}{2}$ & - \\
        & & & & & & & & & \\[-10 pt]
        \hline
        & & & & & & & & & \\[-10 pt]
        $\ket{\phi} \,(\perp \ket{1})$ & $0$ & $\forall \theta$ & - & $0$ & - & $\frac{\pi}{2}$ & $\forall \theta$ & $\frac{\pi}{2}$ & $\frac{\pi}{2}$ \\
        \hline
    \end{tabular}
    \caption{Values of $\theta$ for which various state vectors are orthogonal. We use - to indicate that states are never orthogonal, and $\forall \theta$ to indicate pairs which are always orthogonal. Except for $\ket{\phi}$ these are represented visually in figure \ref{fig:3d_visualisation} The state $\ket{\phi}$ refers to a vector which is constrained to always be mutually orthogonal to $\ket{0}$ and $\ket{\xi}$. Note that the version which is orthogonal to $\ket{1}$ is included in as a row but not a column for space reasons.} 
    \label{tab:orthogonality_relationships}
\end{table}

\section{Effective transverse field Ising Hamiltonian\label{sec:eff_Ising}}

While still assuming that each variable is able to be perfectly restricted to the subspace given in equation \ref{eq:allowed_span}, we can consider the effect of adding Ising type interactions between the $\ket{0}$ and $\ket{1}$ state of each variable, terms of the form
\begin{equation}
    H_{\mathrm{Ising}}=\sum_{j<k}J_{j,k}Z_jZ_k+\sum_j h_jZ_j,
\end{equation}
where $Z=\ketbra{0}{0}-\ketbra{1}{1}$.
For reasons which will become apparent later, we also add a term which offsets the energy of the undefined states and takes the form 
\begin{equation}
    H_{\mathrm{off}}=-\alpha \sum_j\ketbra{u_j}{u_j}. \label{eq:H_off}
\end{equation}
Note that both of these Hamiltonians are diagonal in the expanded computational basis formed by $\ket{u},\ket{0},\ket{1}$. First considering a single variable with a total Hamiltonian taking the form 
\begin{equation}
    H_{\mathrm{tot}}=h Z -\alpha \ketbra{u}{u}.
\end{equation}
Within the basis defined in equation \ref{eq:allowed_span}, $\{\ket{\tilde{+}}=\cos(\theta)\ket{u}+\sin(\theta)\ket{+},\ket{\tilde{-}}=\ket{-}\}$, this Hamiltonian becomes
\begin{equation}
    \left(\begin{array}{cc} -\alpha \cos^2(\theta) & h\sin(\theta)\\ h \sin(\theta) & 0 \end{array}\right).
\end{equation}
Performing a Hadamard transform to enter the $\{\ket{\tilde{0}},
 \ket{\tilde{1}})\}$ basis (defined in equation \ref{eq:zero_one_tilde}) we obtain 
\begin{equation}
    \left(\begin{array}{cc} \frac{\alpha}{2}\cos^2(\theta)+h\sin(\theta) & -\frac{\alpha}{2} \cos^2(\theta) \\ -\frac{\alpha}{2} \cos^2(\theta) & \frac{\alpha}{2}\cos^2(\theta)-h \sin(\theta) \end{array}\right)
\end{equation}
which is effectively the equation of a single Ising qubit subject to a transverse field proportional to $-\alpha\cos(\theta)/2$ and a longitudinal field proportional to $h\sin(\theta)$. Because of the tensor product structure within the construction, this implies that the effect of a $J_{j,k}$ coupling will be proportional to the prefactor on the field term squared and therefore the Hamiltonian of the total system within the $\{\ket{\tilde{0}},\ket{\tilde{1}}\}$ subspace will be
\begin{equation}
    H_{\mathrm{sub}}(\theta)=\frac{\alpha}{2}\cos^2(\theta)\left(n\mathbb{1}-\sum_j\tilde{X}_j\right)+\sin(\theta)\sum_jh_j\tilde{Z}_j+\sin^2(\theta)\sum_{j<k}J_{j,k}\tilde{Z}_j\tilde{Z}_k \label{eq:eff_trans_field}
\end{equation}
where $\tilde{Z}=\ketbra{\tilde{0}}{\tilde{0}}-\ketbra{\tilde{1}}{\tilde{1}}$ and $\tilde{X}=\ketbra{\tilde{0}}{\tilde{1}}+\ketbra{\tilde{1}}{\tilde{0}}$. Equation \ref{eq:eff_trans_field} effectively defines a transverse field Ising model, but where the longitudinal field terms follow a different annealing schedule than the coupling terms. This issue can however be circumvented by replacing fields with couplings to an auxilliary variable which will by construction have the same effective prefactor as coupling terms. Performing this transformation and removing the identity term which only contributes an unmeasurable global phase gives a more canonical form of an effective transverse Ising model
\begin{align}
    H_\mathrm{sub}(\theta)=-A(\theta)\sum_j\tilde{X}_j+B(\theta)\left(\tilde{Z}_a\sum_{j}h_j\tilde{Z}_j+\sum_{j<k}J_{j,k}\tilde{Z}_j\tilde{Z}_k \right),\\
    A(\theta)=\frac{\alpha}{2}\cos^2(\theta),\\
    B(\theta)=\sin^2(\theta).
\end{align}
Within this Hamiltonian the effective spin values are $\average{\tilde{Z}_a\tilde{Z}_j}$. In other words the spin values in the version with fields correspond to whether the spin agrees or disagrees with $a$ in this coupling-only version. One subtle observation to make here is that we have implicitly assumed that projection into the $\{\ket{\tilde{+}},\ket{\tilde{-}}\}$ basis is perfect. More concretely,let us consider a projection is applied in a way which can be described as a matrix in the larger space (either adiabatically through real Hamiltonian elements or dissipatively as can often be effectively described by imaginary Hamiltonian elements) The mathematical assumption in this case is that the elements which forbid the $\ket{\xi(\theta)}$ state are much larger in magnitude than any others. 

We note that importantly, while this implementation effectively realises a transverse field Ising Hamiltonian, it does so without requiring coupling interactions between the variable $Z_iZ_j$ to be tuned over time. In fact, the only variable which needs to be dynamically swept is $\theta$, which is a control parameter on a single variable. While the $h$ and $J$ couplings need to be programmable, this is likely a much less difficult engineering requirement than being able to continuously sweep the values on a desired timescale. Furthermore, many optimisation problems of interest can in fact be mapped using only a single non-zero value of $J$. For example, a maximum cut problem would only require $J_{jk}=1$ for edges which are within the graph and $J_{jk}=0$ otherwise. Likewise for maximum independent set, although in this case additional $h$ terms will also be needed. As a result, highly non-trivial computation can be performed with very limited programmability of the couplings. Several generalisations are also possible in this framework, including biased driving, qudits, and higher order driving terms, these are discussed in section A of the supplemental material.

\section{Zeno incarnations \label{sec:zeno_inc}}

\subsection{Dissipative Zeno effects \label{sec:disp_zeno}}

So far in this manuscript we have assumed that strong coupling is available to implement an unspecified type of Zeno effect to forbid a system from entering specific quantum states. What we now wish to do is to consider a specific implementation, where the Zeno effect is achieved by coupling the states to strong dissipation which would send them into computationally useless states. Some ideas of how such coupling could be implemented have been discussed in \cite{Huang2008zenouniversal,Xu2009ZenoPhase}. We demonstrate in section D of the supplemental material that this implementation allows for a unique kind of satisfiability tester which can determine whether a set of clauses are satisfiable without directly returning the satisfying arrangement.

Within a real system, it is likely that each state which is forbidden in the constraints will couple to it's own ``destroyed'' state rather than a single shared one. Mathematically representing such a system becomes complicated since it would require a master equation which includes transition rates to each destroyed state. However, this picture can be simplified substantially if we treat all such states as the same and as being equivalent and computationally useless, effectively as being a failure of the computation. 

Coupling a state $\ket{\xi}$ to decay in this setting can be effectively represented as acting on the system with 
\begin{equation}
    \exp(-\Gamma t \ketbra{\xi}{\xi} )
\end{equation}
where $\Gamma$ is the (positive) coupling rate to decay. This is equivalent to evolution with an effective anti-Hermitian ``Hamiltonian'', this decay leads to an un-normalised final state with a norm less than or equal to $1$, this final norm represents the probability that a decay has not occurred. A more general model for multiple decay channels is to consider evolution with 
\begin{equation}
    D(t)=\exp(-i\bar{H} t),
\end{equation}
where
\begin{equation}
    \bar{H}=-i\sum_j\Gamma_j\ketbra{\xi_j}{\xi_j}. \label{eq:pure_disap}
\end{equation}
The overline has been added to emphasise that this is an anti-Hermitian matrix and therefore not a ``true'' Hamiltonian. The states $\ket{\xi_j}$ are the set of all (not necessarily orthogonal) states coupled to decay. 

An advantage of this representation is that it allows the same matrix machinery which is typically applied to unitary evolution to be used, since an anti-Hermitian matrix is also diagonalised by a unitary matrix and therefore has a complete set of orthonormal eigenvectors whch are the same operating from the left and right\footnote{Importantly this is not the case in general for complex symmetric matrices, representing systems where both loss and dynamical phases are present meaning that matrix exponentiation cannot simply be performed by exponentiating the eigenvalues and then transforming into a relevant basis.}. Assuming all $\Gamma_j$ are positive, $-i\bar{H}$ will have only non-positive eigenvalues and therefore can only decrease the norm of a state vector, corresponding to decay into computationally useless states.\footnote{This also guarantees that the overall map which we have represented in this way if a density matrix element for dissipated probability is added, with all corresponding off-diagonal elements equal to zero, is completely positive and trace preserving (CPTP). A trace of the initial density matrix which is greater than one would imply either that the overall trace including the dissipated probability is not equal to one, or that the entry corresponding to the dissipated probability is negative.} 

A complication with this method is that while for all $0<\theta\le \pi/2$ the only states which are free from decay have to be those which contain only combinations of $\ket{u}$ and a satisfying assignment, at exactly $\theta=0$ there will be multiple additional states which do not couple to the dissipation. Because the eigenvalues change continuously with $\theta$, these states will be adiabatically connected to states which only couple weakly to dissipation (small magnitude negative eigenvalues) as $\theta$ is increased.

Let us now consider a dissipative implementation of satisfiability problems. Details of the satisfiability implementation are discussed in section C of the supplemental material. The key aspect of this implementation which is important for this discussion is that every clause is implemented by coupling the dissipation to only a single computational basis state. It follows from this construction that any state which is a combination of $\ket{u}$ and $\ket{-}$ where every clause has at least one variable in the $\ket{u}$ configuration will not couple to dissipation at $\theta=0$.

Since hard satisfiability problems tend to have a number of clauses proportional to the number of variables \cite{Kirkpartick994SAT,Xu1999SAT}, there will be an exponentially growing number of such states.  Since it is generally widely suspected that even quantum algorithms will be subject to exponential overheads when solving NP-hard problems, these overheads may still lead to competitive performance. Therefore we cannot automatically rule out the use of the paradigm we have proposed here, but as we show does have consequences on the runtime observed in numerical simulations.

\begin{figure}
    \centering
    \begin{subfigure}[b]{0.475\textwidth}
            \centering
            \includegraphics[width=\textwidth]{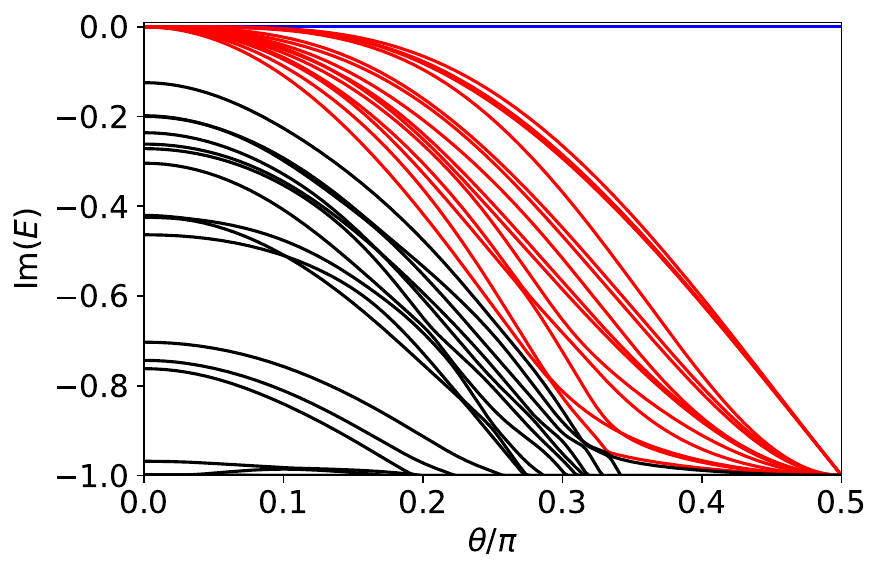}
            \caption[Satisfiable]%
            {Spectrum for satisfiable problem}
            \label{fig:disp_spectrum_sat}
    \end{subfigure}
    \begin{subfigure}[b]{0.475\textwidth}
            \centering
            \includegraphics[width=\textwidth]{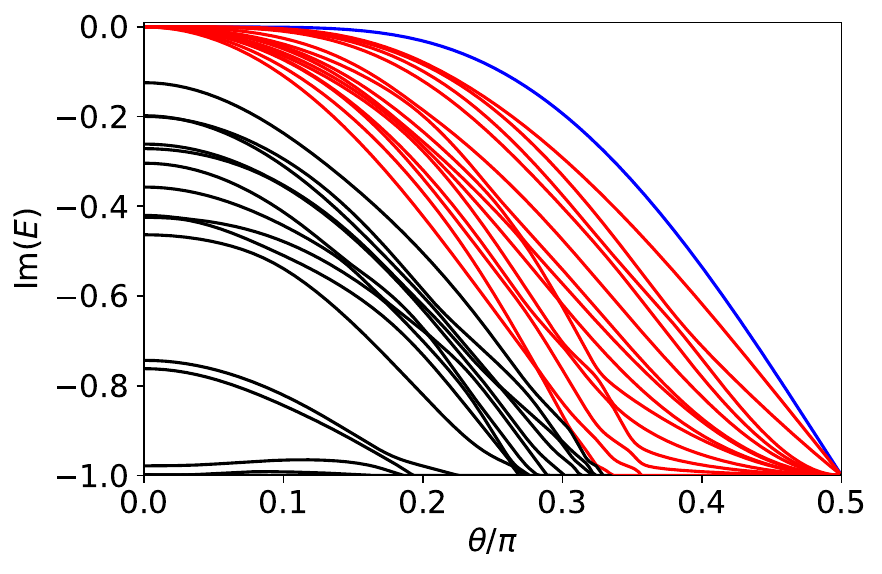}
            \caption[Unsatisfiable]%
            {Spectrum for unsatisfiable problem}
            \label{fig:disp_spectrum_unsat}
     \end{subfigure}
    \caption{Spectrum of $\bar{H}$ defined in equation \ref{eq:pure_disap} for the three satisfiability problem given in section B of the supplemental material and also including equal ($\Gamma_j=1 \forall j$) dissipative $\theta$-dependent coupling to states of the form given in equation \ref{eq:coupling_state}. Subfigure (a) encodes the original problem while (b) encodes the same with an additional clause of $x_0\vee x_1 \vee x_2$, rendering the problem unsatisfiable. Colour coding has been added for ease of reading. The lowest magnitude eigenvalue is coloured blue, while the next $15$, which become degenerate at $\theta=0$ are colour coded red, the rest are coloured black.}
    \label{fig:disp_spectrum}
\end{figure}

To construct a numerical demonstration, we construct a three satisfiability problem with five variables where $\ket{00000}$ is planted as a unique satistfying assignment. This is done by randomly adding clauses with at least one variable is negated until no other satisfying assignment remains. For reproducability, we provide these clauses in section B of the supplemental material. We first examine the $\theta$ dependent spectrum of $\bar{H}$ in the form given in equation \ref{eq:pure_disap}, as well as an unsatisfiable version where the clause $x_0\vee x_1 \vee x_2$ is added. These two spectra can be seen in figure \ref{fig:disp_spectrum}. The key difference between the two, as predicted theoretically is that for a satisfiable problem as depicted in figure \ref{fig:disp_spectrum_sat} a zero eigenvalue eigenstate exists for all $\theta$, while in the unsatisfiable case in figure \ref{fig:disp_spectrum_unsat} this only exists for $\theta=0$ and the magnitude of the smallest magnitude term grows monotonially. We also observe that in both cases there are $15$ additional degenerate eigenstates at $\theta=0$, corresponding to five states with a single $\ket{-}$ tensored with $\ket{u}$ and ten with two $\ket{-}$ terms. It can be verified analytically that such states will not couple to dissipation at $\theta=0$. This degeneracy is lifted for $\theta>0$, but only slowly. As discussed later, this structure has consequences related to the success probabilities of a dissipative protocol.

\begin{figure}
    \centering
    \begin{subfigure}[b]{0.475\textwidth}
        \centering
        \includegraphics[width=\textwidth]{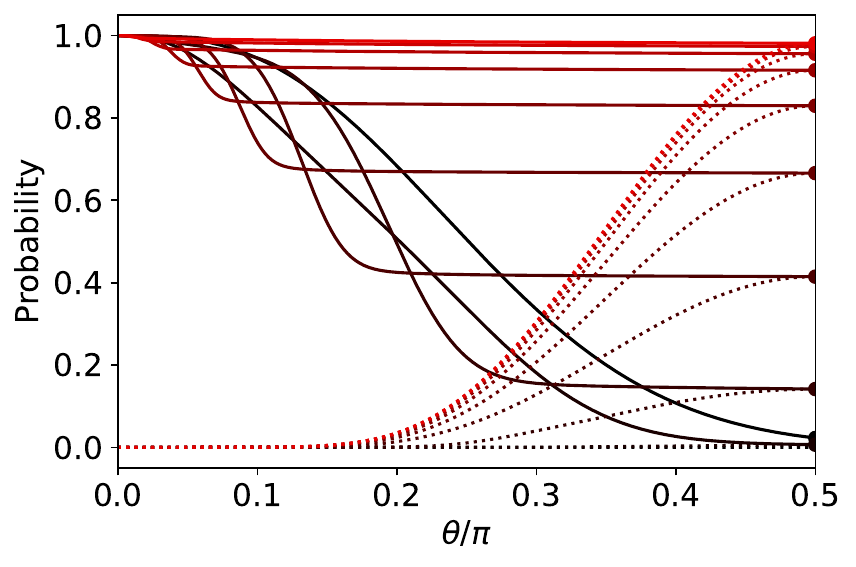}
        \caption[]%
            {Intermediate $\theta$ values}
            \label{fig:sat_disp_norm_success_theta}
    \end{subfigure}
    \begin{subfigure}[b]{0.475\textwidth}
        \centering
        \includegraphics[width=\textwidth]{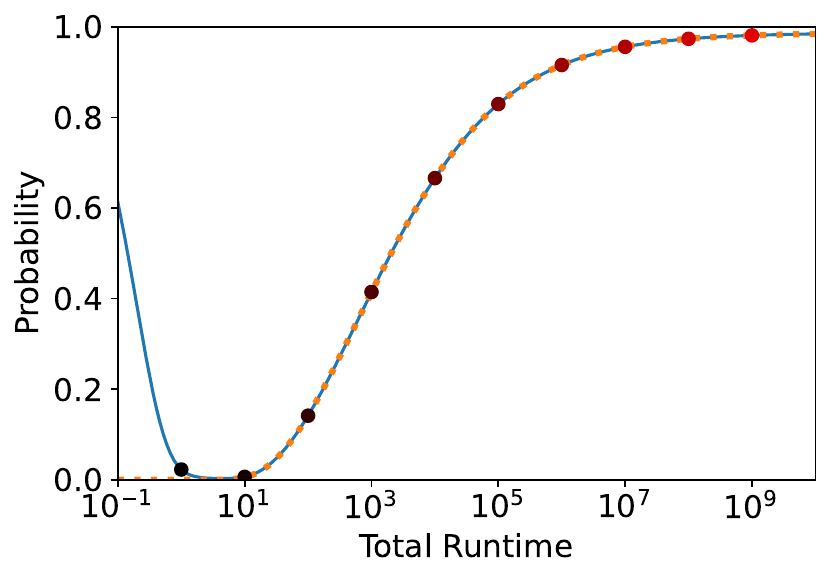}
        \caption[Satisfiable]%
            {At final time}
            \label{fig:sat_disp_norm_success_time}
    \end{subfigure}
    \caption{Probability that dissipation has not happened (solid lines), and probability of finding the satisfying assigment (dashed lines). This plot shows the results of a linear ramp of $\theta$ for a total runtime shown in subfigure (b). This problem uses an $\bar{H}$ defined in equation \ref{eq:pure_disap} for the three satisfiability problem given in section B of the supplemental material and also including equal ($\Gamma_j=1 \forall j$) dissipative $\theta$-dependent coupling of to states of the form given in equation \ref{eq:coupling_state}. Subfigure (a) shows the time dependent quantities, with total runtime colour coded as the circles in subfigure (b). Note the logarithmic scale spanning $11$ orders of magnitude in subfigure (b).}
    \label{fig:sat_disp_norm_success}
\end{figure}

From the spectrum in figure \ref{fig:disp_spectrum_unsat}, we can deduce that even for the eigenstate with the lowest magnitude eigenvalue, there will be rapid dissipation, leading to a very small amplitude in this formalism. On the other hand, the zero eigenvalue which can be observed in figure \ref{fig:disp_spectrum_sat} confirms that non-trivial solving behaviour is possible. As figure \ref{fig:sat_disp_norm_success}, shows the probability to find a satisfying assignment does indeed approach one as runtime is increased. We note however from figure \ref{fig:sat_disp_norm_success} that this approach is very slow, with modest success probability appearing in with runtimes values in the tens, but close appraoch to unity not occuring until $10^9$. Such a slow approach is likely the result of very weak Zeno effects due to eigenstates which only couple weakly to the dissipation at low $\theta$. A fact confirmed by figure \ref{fig:sat_disp_norm_success_theta}, which shows dissipation occurring increasingly early in the sweep as runtimes are increased. While beyond the scope of our current work, this also suggests that success probability could be improved by performing a non-uniform sweep, which proceeded slowly at small $\theta$ values. 

We now observe that a state of the form
\begin{align}
    \ket{\phi(\theta)}=\frac{1}{\left(1+\sin^2(\theta)\left(\sqrt{2}-1\right)^2\right)^{\frac{n}{2}}}\bigotimes_{i=1}^{n} \left(\cos(\theta) \ket{u}+\sqrt{2}\sin(\theta) \ket{s_i}\right) \label{eq:sat_state_s}\\
    =\frac{1}{\left(1+\sin^2(\theta)\right)^{\frac{n}{2}}}\bigotimes_{i=1}^{n}  \left(\cos(\theta) \ket{u}+\sin(\theta)\left(\ket{+}+(-1)^{s_i}\ket{-}\right) \right)\label{eq:sat_state_pm} \\
    =\frac{1}{\left(1+\sin^2(\theta)\right)^{\frac{n}{2}}}\bigotimes_{i=1}^{n} \left(\ket{\tilde{+}}+(-1)^{s_i}\sin(\theta)\ket{\tilde{-}}\right) \\
    =\frac{1}{\left(2+2\sin^2(\theta)\right)^{\frac{n}{2}}}\bigotimes_{i=1}^{n}\left((1+(-1)^{s_i}\sin(\theta))\ket{\tilde{0}}+(1-(-1)^{s_i}\sin(\theta))\ket{\tilde{1}}\right),
\end{align}
 where $s_{i}$ is an element within a satisfying assignment, will not couple to dissipation.

 The dissipative implementation does have an interesting property; if a satisfying arrangement $s$ exists, then the construction in equation \ref{eq:sat_state_s} would correspond to a state which does not overlap any states coupled to decay and therefore will not decay itself, for all values of $\theta$. On the other hand, if no satisfying arrangement exists, then for any $0<\theta \le \pi/2$, any state will have to decay. Therefore if we initialize in $\sum_j \ket{u_j}$ and perform a Zeno sweep $\theta$ to any value greater than zero, then if a satisfying clause does not exist, this state will be unstable and eventually decay. Since $\theta$ can be substantially less than $\pi/2$, the measured state will be substantially different from the actual satisfying arrangement, and therefore we have a method using a small sweep to test if an expression has a satisfying arrangement without directly finding the arrangement itself. Such a detection tool could potentially find use in constructing a complete solver in proving unsatisfiability. Although it is worth noting the implicit assumption that the system evolves slowly enough to remain in the relvant subspace, which will be difficult in practice.

In principle such a solver could be used iteratively since any $\{ \ket{0}, \ket{1} \}$  values which were obtained when measuring a state out are guaranteed to belong to a satisfying assignment. One could then fix these bit values and solve a smaller satisfiability problem using the same method until all bits are assigned. While such an iterative method is intriguing and represents a new way to solve satsifiability problems, a detailed study is beyond the scope of our current work. A brief outline of how such an algorithm could be constructed appears in section D of the supplemental material.

\subsection{Repeated Measurements}

\begin{figure}
    \centering
    \begin{subfigure}[b]{0.475\textwidth}
        \centering
        \includegraphics[width=\textwidth]{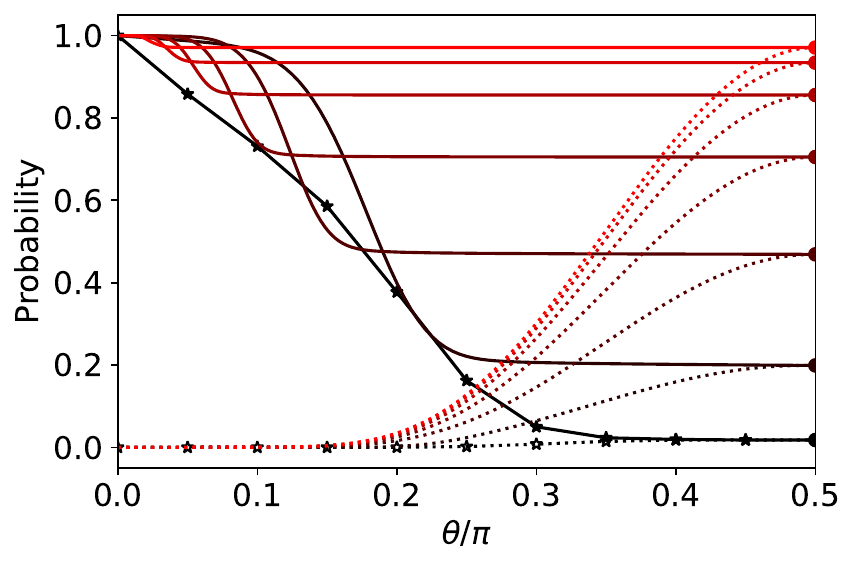}
        \caption[]%
            {Intermediate $\theta$ values}
            \label{fig:sat_meas_norm_success_theta}
    \end{subfigure}
    \begin{subfigure}[b]{0.475\textwidth}
        \centering
        \includegraphics[width=\textwidth]{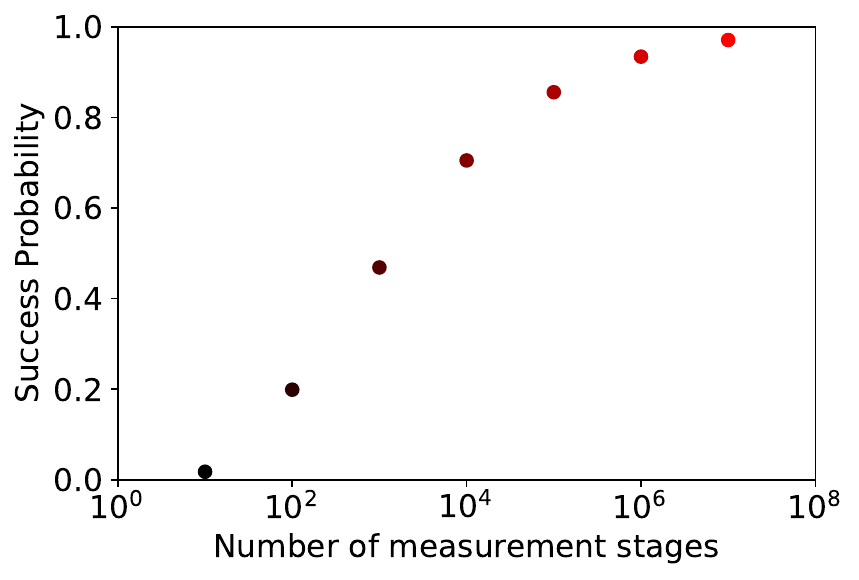}
        \caption[Satisfiable]%
            {After all measurements}
            \label{fig:sat_meas_norm_success_time}
    \end{subfigure}
    \caption{ This plot shows the results of successive measurements using a linear ramp of $\theta$ for a total number of measurements shown in (b). Measurements are performed of whether the system are in a set of states $\ket{\xi}$ consisting of $\ket{\xi(\theta)}$ from equation \ref{eq:coupling_state} followed by a measurement of whether the system is in any states which violate clauses of the three satisfiability problem defined in section B of the supplemental material. Subfigure (a) depicts the probabilities as a function of $\theta$ with total number of measurements colour coded as the circles in subfigure (b). In subfigure (a), probability that a $\ket{\xi}$ state has not been measured (solid lines), and probability of finding the satisfying assignment (dashed lines). For the lowest number of measurements, ten measurements, the individual measurements are highlighted as stars, with lines acting as a guide to the eye. Subfigure (b) shows success probability, which is by definition equal to the probability that no $\ket{\xi}$ state has been measured. Note the logarithmic scale in subfigure (b).}
    \label{fig:sat_meas_norm_success}
\end{figure}

An alternative implementation is to perform repeated measurements of the $\ket{\xi}$ states, while slowly modifying $\theta$. Since this is effectively a digitised version of the dissipative implementation, with dissipation occurring at discrete steps in the form of measurements, one may expect qualitatively similar behaviour, especially for a large number of measurements. Namely we expect that the probability of having never measured a forbidden state $\ket{\xi}$ would drop off at increasingly low values of $\theta$ the more measurements are taken. As figure \ref{fig:sat_meas_norm_success} shows, this intuition is indeed correct, with figure \ref{fig:sat_meas_norm_success_theta} showing the same characteristic drop in probability that no measurement has ``failed'' and figure \ref{fig:sat_meas_norm_success_time} showing the same slow growth in success probability. It is worth remarking that solving satisfiability problems using only measurements has been considered in other works, but using different, qubit based methods \cite{Benjamin2017MeasurementSAT,Zhang2024MeasrurementSAT} relying on repeated projective measurements in a non-orthogonal basis.

\subsection{Adiabatically enforced satisfiability constraints}

\begin{figure}
    \centering
    \begin{subfigure}[b]{0.475\textwidth}
            \centering
            \includegraphics[width=\textwidth]{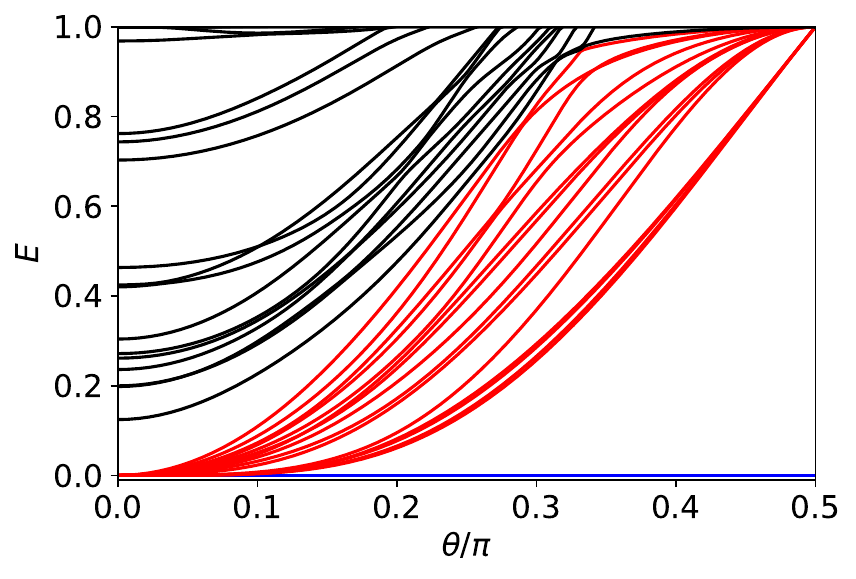}
            \caption[no offset]%
            {Spectrum without offset}
            \label{fig:dadiabat_no_off_spectrum_sat}
    \end{subfigure}
    \begin{subfigure}[b]{0.475\textwidth}
            \centering
            \includegraphics[width=\textwidth]{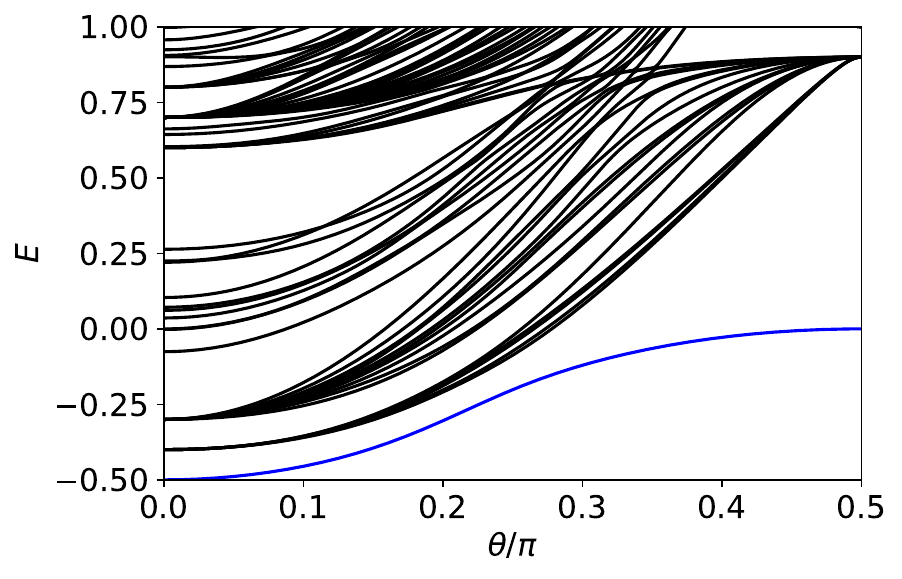}
            \caption[offset]%
            {Spectrum with offset}
            \label{fig:adiabat_0p1_off_spectrum_sat}
     \end{subfigure}
    \caption{Spectrum of $H$ defined in equation \ref{eq:adiabat_nofrust} for the three satisfiability problem given in section B of the supplemental material and also including equal ($c_j=1 \forall j$) $\theta$-dependent coupling of to states of the form given in equation \ref{eq:coupling_state}. Subfigure (a) does not contain an offset (b) includes an offset of the form given in equation \ref{eq:H_off} with $\alpha=0.1$. Colour coding has been added for ease of reading, in both the lowest magnitude eigenvalue is coloured blue, while in (a) the next $15$, which become degenerate at $\theta=0$ are colour coded red, the rest are coloured black.}
    \label{fig:adiabat_spectrum}
\end{figure}

Next we consider applying these methods within an adiabatic setting. An advantage of this setting is that we are not restricted to a state with a zero eigenvalue. In other words, we can take advantage of frustration, where competing terms can be used to construct a more complicated ground state. Before discussing this however, we first consider the same implementation explored in section \ref{sec:disp_zeno} but implemented adiabatically, based on a real Hamiltonian
\begin{equation}
    H_\mathrm{proj}=\sum_jc_j\ketbra{\xi_j}{\xi_j}. \label{eq:adiabat_nofrust}
\end{equation}
A non-zero eigenvalue will not lead to dissipation in the adiabatic case, it is natural to also consider a frustrated version. For this reason, it is possible to add an offset Hamitonian of the form given in equation \ref{eq:H_off}. We add such a term with an offset strength of $\alpha=0.1$. The resulting spectra can be seen in figure \ref{fig:adiabat_spectrum}. We first note that by definition figure \ref{fig:dadiabat_no_off_spectrum_sat} is simply an inverted version of figure \ref{fig:sat_disp_norm_success_theta}, including the degeneracy at $\theta=0$. On the other hand, when an offset is added in figure \ref{fig:adiabat_0p1_off_spectrum_sat}, the initial degeneracy disappears leaving a Hamiltonian with a finite gap at all times. Note however that the lowest eigenvalue is no longer $0$ except for at $\theta=\pi/2$. As a consequence, a direct implementation of a dissipative analog of this Hamiltonian would simply result in decay. The adiabatic analogy to decay rate is an unmeasurable global phase, which is not detrimental to (in fact has no effect on) the final success probability.

\begin{figure}
\centering
    \begin{subfigure}[b]{0.475\textwidth}
        \centering
        \includegraphics[width=\textwidth]{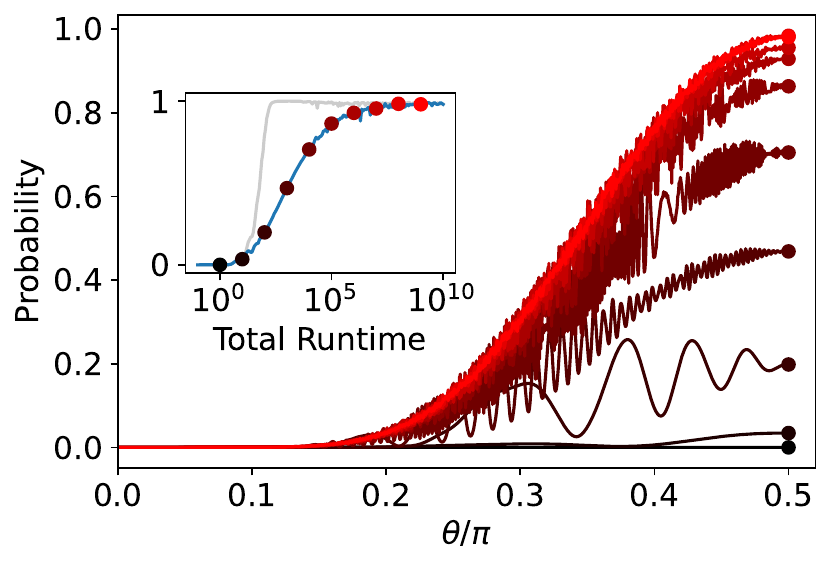}
        \caption[]%
            {No offset}
            \label{fig:sat_adiabat_no_offset_success_theta}
    \end{subfigure}
    \begin{subfigure}[b]{0.475\textwidth}
        \centering
        \includegraphics[width=\textwidth]{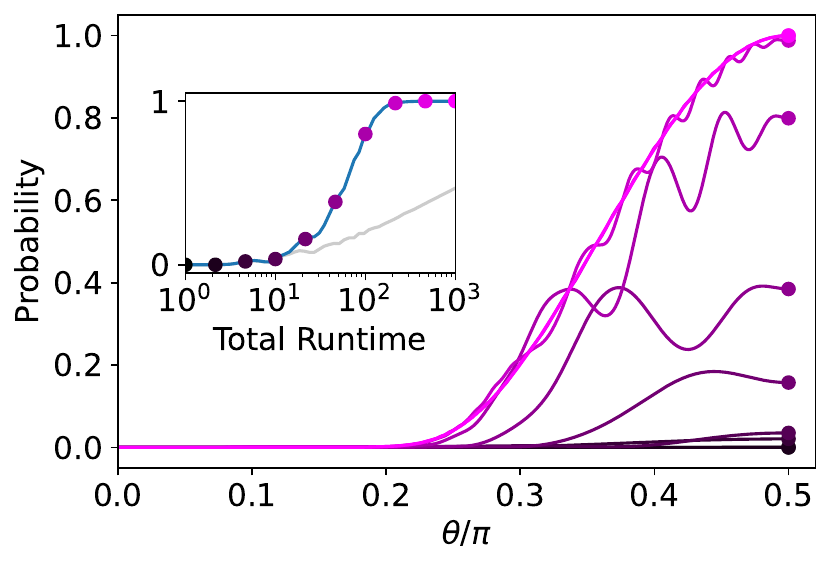}
        \caption[Satisfiable]%
            {Added offset}
            \label{fig:sat_adiabat_0p1_offset_success_theta}
    \end{subfigure}
    \caption{Probability of finding the satisfying assignment This plot shows the results of a linear ramp of $\theta$ for a total runtime shown in the insets using colour coding to match the curves to the points. This problem uses an $H$ defined in equation \ref{eq:adiabat_nofrust} for the three satisfiability problem given in section B of the supplemental material and also including equal ($c_j=1 \forall j$) $\theta$-dependent coupling of states of the form given in equation \ref{eq:coupling_state}. Subfigure (a) shows the results with no offset, the adiabatic analogy of figure \ref{fig:sat_disp_norm_success} subfigure (b) shows the result for the same system with an additional offset of the form given in equation \ref{eq:H_off} with a strength of $\alpha=0.1$. Note the logarithmic scale in both insets, and particularly that the scale in (a) spans $11$ orders of magnitude. Within both subfigures, the full colour curve shows the final success probabilities of that implementation, while the gray curve shows the other for comparison; so the gray curve in (a) matches the full colour curve in (b) and vice-versa.}
    \label{fig:sat_adiabat_success}
\end{figure}

A natural question to ask is what effect this additional offset has on the dynamics. As figure \ref{fig:sat_adiabat_success} shows, the additional gap caused by the offset causes the success probability as function of runtime to approach one much faster. While figure \ref{fig:sat_adiabat_no_offset_success_theta} shows very similar final success probabilities to figure \ref{fig:sat_disp_norm_success}, adding an offset as shown in figure \ref{fig:sat_adiabat_0p1_offset_success_theta} leads to a steep rise in success probability. 

\section{Non-perturbative Constraints\label{sec:adiabat_num}}

While the primary goal of the paradigm we have developed of projecting out states has been to explore alternative Zeno effects to traditional adiabatic methods, these methods also make an important contribution to what can be achieved adiabatically. In particular, we will demonstrate that non-trivial computation can be performed even when infinitely-strong constraints are applied, this is backed up by calculations performed in section E of the supplemental material. This directly contrasts to the way in which constraints are usually implemented in an adiabatic setting, by applying very strong Ising terms, which render the dynamics within the valid subspace perturbative unless it is connected by single bit flips. In other words, these require the system to travel through high energy states to traverse the state space, the rate of these transitions scales inversely with the penalty strength creating a tradeoff. With the three-state implementation, no such tradeoff exists. As a numerical demonstration, we consider a simple problem of single-body Ising terms acting on $5$ variables under a three-hot constraint. This constraint adds $c_j=1$ for all bitstrings which cannot satisfy the constraint regardless of the values assigned to an $\ket{u}$ states that may be present. We randomly selected the five single body Ising terms such that three are negative and two are positive
\begin{equation}
\vec{h}=\left[ -0.67513783, -0.62099006, -0.14675767,  0.72688415,  0.56602992\right]. \label{eq:rand_h}
\end{equation}
As a result, the unconstrained solution state will be $\ket{00011}$, while the constrained one will be $\ket{00111}$. 

\begin{figure}
\centering
    \begin{subfigure}[b]{0.475\textwidth}
        \centering
        \includegraphics[width=\textwidth]{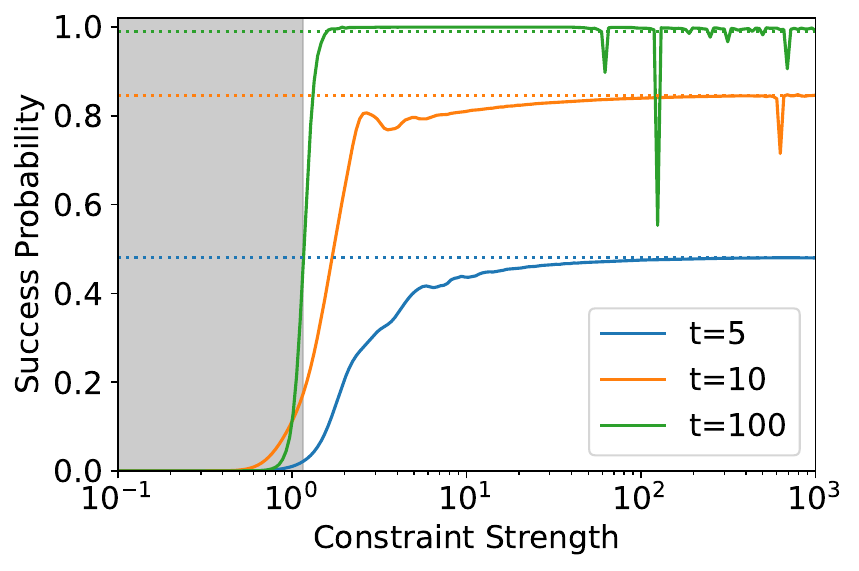}
        \caption[]%
            {Three-state implementation}
            \label{fig:fields_3hot_h5_adiabat_3}
    \end{subfigure}
    \begin{subfigure}[b]{0.475\textwidth}
        \centering
        \includegraphics[width=\textwidth]{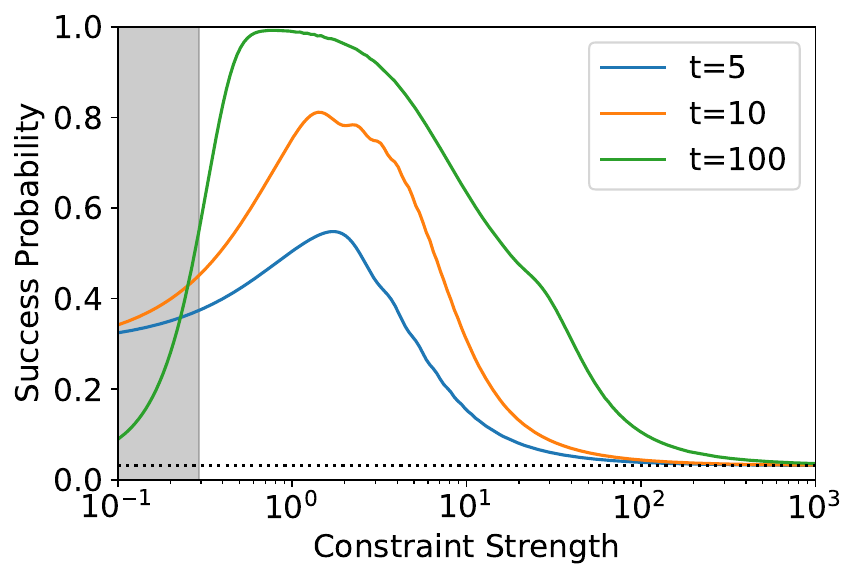}
        \caption[]%
            {Transverse-field implementation}
            \label{fig:fields_3hot_h5_adiabat_tf}
    \end{subfigure}
    \caption{Success probability versus constraint strength for a system of single-body terms (randomly selected, given in equation \ref{eq:rand_h}) subject to a three-hot constraint implemented by giving all configurations which violate the constraint an additional energy equal to the constraint strength. $t$ is the total runtime of the algorithm. (a) Three state Zeno implementation defined in equation \ref{eq:adiabat_nofrust} using linear sweep in $\theta$, with the $c_i$ terms corresponding to $\theta$ dependent forbidden states also equal to the constraint strength and offset of $\alpha=1$ (defined in equation \ref{eq:H_off}). (b) Transverse field implementation using a linear sweep in $s$ with the Hamiltonian defined in equation \ref{eq:tf_lin_sweep}. The dotted lines in (a) indicate the results of perfect projection into the allowed subspace, while the dotted line in (b) is the success probability from random guessing. The grey area indicates constraint strengths which are not strong enough for the final ground state to correspond to the correct solution. The sharp dips in plot (a) are likely artifacts of the numerical matrix diagonalisation algorithm arising due to the matrix being highly ill-conditioned.}
    \label{fig:fields_3hot_h5_adiabat}
\end{figure}

We compare this to the same constraint applied to the same problem, but in a traditional transverse field setting, where each variable can only take values of $\{\ket{0},\ket{1}\}$, and a transverse field is linearly swept from $s=0$ to $s=1$ over a time $t$
\begin{equation}
H(s)=-(1-s)\sum_jX_j+s H_\mathrm{problem}. \label{eq:tf_lin_sweep}
\end{equation}
We initiate this system in its ground state, which is an equal positive superposition of all computational basis states. Figure \ref{fig:fields_3hot_h5_adiabat} compares the results of these two strategies at different constraint strengths. As we can see from figure \ref{fig:fields_3hot_h5_adiabat_3}, strengthening the constraints causes an approach to behaving as a perfectly projected system, with a finite success probability comparable to the best success probability seen in figure \ref{fig:fields_3hot_h5_adiabat_tf} for the same runtime. On the other hand in figure \ref{fig:fields_3hot_h5_adiabat_tf} there is a tradeoff between constraint strength and realising dynamics which are able to effectively solve the problem, leading to success probabilities no better than random guessing as constraints become very strong. 

\section{Experimental Implementation\label{sec:exp_imp}}

A key advantage to the methods proposed here is that in our proposal all dynamics are achieved by sweeping a term which acts only on a single three-state system, coupling between terms do not need to be continuously tuned over time. Purely dissipation-based implementations have the disadvantage of not being able to directly implement frustration. A solution to this problem which is used in optical Ising machines is to include both loss and gain, a strategy which has proven fruitful and has been extensively discussed elsewhere in the literature, \cite{Kumar2020opticalIsing,Yamamoto2017CoherentIsing,Yamamoto2020CoherentIsing,Lu2023CoherentIsing} and where substantial opportunities exist. This direction is particularly timely because implementations of Zeno blockade effects have been demonstrated in non-linear optical systems \cite{Chen2017ZenoChip}. In addition to potential use in combinatorial optimisation, these effects have been shown to be useful in performing interaction free switching \cite{Huang2010switching,McCusker2013switching}, entangled photon generation \cite{Huang2012entangled}, and quantum gates \cite{Huang2012Fredkin,Sun2013ZenoGates}.

 Settings which are also interesting are those with driven transitions between a small number of energy levels. Such structures occur naturally in both atomic systems and transmon qubits are a natural setting for implementing systems with more than two variables. Atomic systems naturally have multiple energy levels, while transmon\footnote{It is not true that all superconducting qubit architectures are naturally amenable to extending to higher than binary quantum systems, flux qubits for example have a natural double-well structure which cannot be naturally extended to higher-than-binary variables \cite{Krantz2019Superconducting}.} systems make energy level transitions individually addressable by using Josephson junctions, introducing anharmonicity to an otherwise harmonic superconducting LC circuit \cite{Krantz2019Superconducting,Roth2023Transmon}. 

\begin{figure}
    \centering
    \includegraphics[width=10 cm]{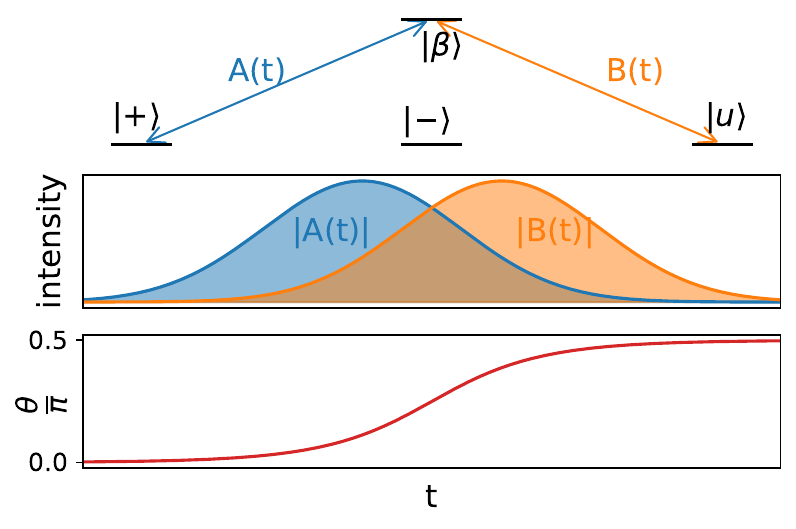}
    \caption{Cartoon representation of a STIRAP-like implementation of the driving protocol. The top frame shows the two time dependent couplings from equation \ref{eq:H_STI}. The mid plot shows a time dependent protocol consisting of the application of both pulses. The value of $\theta$ as defined in equation \ref{eq:theta_STI} is continuously swept as the pulses are applied as is illustrated in the bottom plot.}
    \label{fig:STIRAP_protocol_cartoon}
\end{figure}

In both settings the basic ingredients which are available are the same; essentially electromagnetic waves (typically laser light in an atomic setting and microwaves in a superconducting setting) are used to mediate transitions between states. At the level of detail which we are discussing it is therefore not necessary to distinguish between the two settings. If we consider the situation where an additional state $\ket{\beta}$ is available, then the task of coupling to $\ket{\xi}$ while maintaining $\ket{\tilde{+}}$ as a dark state can be achieved by a protocol reminiscent of the celebrated stimulated Raman adiabatic passage (STIRAP) protocol\cite{Vitnov2017STIRAP}. Concretely, this protocol would consist of applying a time dependent Hamiltonian of the form\footnote{the conventional STIRAP Hamiltonian does not have a $-$ sign in front of the $B$ term, this arises from earlier definitions and could be removed without changing other results by a definition change $\ket{u}\rightarrow -\ket{u}$} 
\begin{equation}
    H_\mathrm{STR}(t)=A(t)\left(\ketbra{+}{\beta}+\ketbra{\beta}{+}\right)-B(t)\left(\ketbra{u}{\beta}+\ketbra{\beta}{u}\right). \label{eq:H_STI}
\end{equation}
We now observe that if we set
\begin{equation}
\theta=\tan^{-1}\left(\frac{B(t)}{A(t)}\right), \label{eq:theta_STI}
\end{equation}
then $\ket{\tilde{+}}$ will be an eigenstate of equation \ref{eq:H_STI} with an eigenvalue of $0$, effectively a ``dark state'' since it does not couple at all to any of the Hamiltonian terms. The other two eigenstates will be $1/\sqrt{2}(\ket{\beta}\pm\ket{\xi})$ with eigenvalues $\pm \sqrt{A^2(t)+B^2(t)}$. By symmetry only $1/\sqrt{2}(\ket{\beta}+\ket{\xi})$ will be populated, so therefore this system effectively applies an additional phase to $\ket{\xi}$ while leaving $\ket{\tilde{+}}$ and $\ket{-}$ unchanged. A protocol for applying pulses driving these transitions which effectively sweeps $\theta$ is represented visually in figure \ref{fig:STIRAP_protocol_cartoon}. 

Many additional techniques which are known in the STIRAP literature could also be applied here, in particular, as is usually the case in conventional STIRAP, the state $\ket{\beta}$ could decay, allowing for implementation of dissipative Zeno effects. Similarly, if discrete pulses are applied to transfer amplitude into $\ket{\beta}$ and a measurement which is only able to discern if the system is in $\ket{\beta}$ or not is performed, then this method could apply a Zeno effect which is instantiated through measurement.

\subsection{Errors}

A potential concern for this proposal is that a three (or four) level system may be more prone to errors than a more simple two level system. While the details of such a discussion would depend on the exact nature of an implementation and therefore beyond the scope of this discussion, it is worth making some general remarks. The first remark is that while each variable is comprised of more than two levels, the subspace in which the dynamics are confined is still only comprised of two levels. Moreover, in many situations, single qubit driving can generally be made very strong, much stronger than two qubit coupling (for example in an atomic system by using a high level of laser power). In fact expanding a Hilbert space to incorporate additional energy levels is a fairly standard technique in engineering of atomic systems. The STIRAP protocol which we previously mentioned is one example, in which high fidelity transfer between two states is mediated by a third (potentially very noisy) state \cite{Vitnov2017STIRAP}. A similar concept is Rydberg dressing, where a low energy atomic state is coupled to ``dressed'' by coupling it to a Rydberg state to mediate strong interactions between atoms. In fact Rydberg dressing has been proposed as a way to engineer high-fidelity quantum gates \cite{Keating2015Rydberg}. As another example, Holonomic gates will always need coupling to an additional energy level to be achieved \cite{Zanardi1999Holonomic,Kodai2018Holonomic}, but are recognized as a promising path toward high-fidelity quantum computation.

The engineering of individual variables are more complex than in a situation where transverse driving is directly implemented between two levels. While this may not necessarily increase the error, added complexity should be justified somehow as a general design principle of of any system. In this case we believe such justification may exist, namely through the ability to implement constraints non-perturbatively, as discussed in section \ref{sec:adiabat_num}.  

\section{Numerical Methods \label{sec:num_meth}}

All numerical simulations were performed in Python 3. All sweeps involved $1000$ evenly spaced time steps, which was found to be sufficiently well converged to make accurate plots. Projection into valid subspaces was performed by numerically diagonlising and keeping only eigenvectors where the magnitude of the eigenvalues is less than $10^{-9}$. Measurements were simulated by sequentially applying projection matrices by building projection matrices and performing matrix multiplication routines. All times are in dimensionless units with scales determined by the energies.

In addition to \textbf{numpy} \cite{numpy} and \textbf{scipy} \cite{2020SciPy-NMeth}, the \textbf{matplotlib} \cite{hunter2007matplotlib} package was heavily used in making plots. No high performance computing resources were used in the creation of the plots, but some plots did require more than a day of compute time on a standard laptop.

For reproducbility purposes the code to produce the plots has been uploaded to the QCi public github and can be found at \url{https://github.com/qci-github/eqc-studies/tree/main/Zeno_constraint}.

\section{Discussion and Conclusions \label{sec:discuss_conclude}}

Within this work we have identified a number of opportunities and a key challenge to generalising Zeno-effect computing beyond the usual adiabatic setting. The opportunities include a method for developing solvers which can be used in a purely dissipative setting or in a setting where no operation other than measurements are performed unlike measurement based quantum computation. Our method allows the system to be initialised in a simple product state, rather than a complex cluster state. Furthermore, these systems have the interesting property of being able to test if a set of clauses is satisfiable without obtaining the full solution, raising the possibility of developing complete rather than incomplete satisfiability solvers. 

The key challenge we observe is that the mathematical structure of the systems we develop lead to weak coupling and correspondingly weak Zeno effects, which manifest as a slow increase in success probability with increased total runtime. This suggests that systems including both loss and gain, or supporting frustration, may fundamentally be more powerful solvers. This naturally points toward the power of two existing paradigms (in addition to the emerging paradigm of entropy computing, which acted as motivation to carry out this work \cite{nguyen2024entropycomputing}). Firstly a paradigm based on encoding optimality in phase, since phases naturally support frustration, this route is typified by the paradigm of adiabatic quantum computing, and it's natural extensions into annealing. A second paradigm is a fundamentally many-body one where both gain and loss are present, and is typified by coherent Ising machines and entropy computing. An interesting question which is far beyond the scope of this paper is whether other paradigms exist which can overcome this challenge.

\section*{Acknowledgments}

All authors were completely supported by Quantum Computing Inc.~in the completion of this work and would like to thank Uchenna Chukwu for a critical reading of the paper and useful comments.

\ifx \justbeingincluded \undefined

\bibliographystyle{unsrt}
\bibliography{references}  

\end{document}

\fi

\ifx \justbeingincluded \undefined

\documentclass[english]{article}
\usepackage[T1]{fontenc}
\usepackage[latin9]{inputenc}
\usepackage{babel}
\usepackage{graphicx,amsmath,amssymb,color}
\usepackage[normalem]{ulem}
\usepackage{amsfonts}
\usepackage[toc,page]{appendix}
\usepackage{hyperref}
\usepackage{latexsym}
\usepackage{amsfonts}
\usepackage{algpseudocode}
\usepackage{amsthm}
\usepackage{mathrsfs}
\usepackage{color,verbatim}
\usepackage{psfrag}

\usepackage{rotating} 
\usepackage{bbold} 
\usepackage{multirow}

\usepackage{subcaption} 

\usepackage[svgnames]{xcolor}
\usepackage{tikz}

\usepackage{algorithm}

\newcommand{\bra}[1]{\langle#1 |}
\newcommand{\ket}[1]{|#1 \rangle}
\newcommand{\bigket}[1]{\Bigl \lvert#1  \Bigr \rangle}
\newcommand{\braket}[2]{\left \langle #1 \middle \vert #2 \right \rangle}
\newcommand{\bigbraket}[2]{\left \langle #1 \middle\vert #2 \right \rangle}
\newcommand{\ketbra}[2]{\vert #1 \rangle \! \langle #2 \vert}
\newcommand{\overlap}[2]{\left \langle #1 | #2 \right\rangle}
\newcommand{\average}[1]{\langle #1 \rangle}
\newcommand{\sandwich}[3]{\left \langle #1 \middle \vert #2 \middle \vert #3 \right\rangle}
\newcommand{\innerprod}[2]{\left \langle #1 | #2 \right\rangle}
\newcommand{\nick}[1]{\textcolor{blue}{Nick: #1}}
\definecolor{myDarkRed}{rgb}{0.5, 0, 0} 
\newcommand{\catherine}[1]{\textcolor{myDarkRed}{Catherine: #1}}
\newcommand{\uchenna}[1]{\textcolor{magenta}{Uchenna: #1}}
\definecolor{myOrange}{rgb}{0.7, 0.3, 0} 
\definecolor{myDarkGreen}{rgb}{0, 0.7, 0} 
\newcommand{\raouf}[1]{\textcolor{myDarkGreen}{Raouf: #1}}
\definecolor{myAqua}{rgb}{0, 0.3, 0.3}
\newcommand{\jesse}[1]{\textcolor{myOrange}{Jesse: #1}}
\newcommand{\removed}[1]{\textcolor{red}{\sout{#1}}}
\newcommand{\added}[1]{\textcolor{blue}{#1}}

\title{Supplement to: Zeno-effect Computation: Opportunities and Challenges}
\author{Jesse Berwald$^\dagger$, Nicholas Chancellor, Raouf Dridi$^\dagger$}
\date{Quantum Computing Inc (QCi), 5 Marine View Plaza Hoboken NJ 07030 USA \newline
$\dagger$ current affiliation: Qamia Quantum Technologies
Albuquerque, USA and Abu Dhabi, UAE \\ \today}

\begin{document}

\maketitle

\tableofcontents

\appendix
\else

\appendix
\addcontentsline{toc}{section}{Supplemental Material}
\part*{Supplemental Material}

\fi

\section{Generalisations of transverse Ising model \label{app:TI_gen}}

\subsection{Biased driving}

Being able to bias an annealing search has been a topic of recent interest given the large space of hybrid quantum/classical algorithms available \cite{chancellor17b,Callison2022hybrid}. Various methods have been proposed, including different forms of reverse annealing\footnote{Reverse annealing is a somewhat ambiguous term, since the same name has been applied to a number of distinct protocols. See \cite{Callison2022hybrid} for details.} \cite{perdomo-ortiz11,chancellor17b,Ohkuwa2018Reverse,Yamashiro2019Reverse,Callison2022hybrid}. An alternative method is to use driving fields which are not completely transverse, but contain some Pauli $Z$ character \cite{Duan2013a,Grass19a}. An advantage of these methods is that unlike reverse annealing they are compatible with energy redistribution arguments \cite{Callison21a}, which may explain numerically-observed high success probabilities for the adiabatic limit in rapid quenches. The versions discussed here however will effectively induce a Hamiltonian of a more complex form which is not compatible with these arguments.

We find that a novel type of biased search can be implemented in our system by coupling to 
\begin{equation}
    \ket{\xi(\theta)}=-\sin(\theta)\ket{u}+\cos(\theta)\ket{\zeta_+(\phi)}, \label{eq:coupling_state_bias}
\end{equation}
where
\begin{equation}
    \ket{\zeta_+(\phi)}=\cos(\phi)\ket{0}+\sin(\phi)\ket{1}.
\end{equation}
If we further define 
    \begin{equation}
    \ket{\zeta_-(\phi)}=\sin(\phi)\ket{0}-\cos(\phi)\ket{1},
\end{equation}
then we have 
\begin{align}
 \{\ket{\tilde{0}}=\cos(\phi)\left[\cos(\theta)\ket{u}+\sin(\theta)\ket{\zeta_+(\phi)}\right]+\sin(\phi)\ket{\zeta_-(\phi)},\nonumber \\
 \ket{\tilde{1}}=\sin(\phi)\left[\cos(\theta)\ket{u}+\sin(\theta)\ket{\zeta_+(\phi)}\right]-\cos(\phi)\ket{\zeta_-(\phi)}\}.   \label{eq:zero_one_tilde_bias}
\end{align}
For $\phi=\pi/4$, this reduces to unbiased search, whereas $0<\phi<\pi/4$ corresponds to a bias toward $\ket{0}$, and $\pi/4<\phi<\pi/2$ corresponds to a bias toward $\ket{1}$, with increasing bias the closer the angle gets to $0$ or $\pi/2$, respectively.

We can see a key aspect of the structure of the biased search, the initial $\ket{u}$ state starts as an unequal superposition of $\ket{\tilde{0}}$ and $\ket{\tilde{1}}$. Applying $-\alpha \ketbra{u}{u}$ yields terms of the form, 
\begin{equation}
    -\alpha\cos^2(\theta) \left(\mathbb{1}+\left(\cos^2(\phi)-\sin^2(\phi) \right)\tilde{Z}+\left( \cos(\phi)\sin(\phi) \right)\tilde{X}\right).
\end{equation}

Unfortunately, an additional effect of working in the $\{\ket{\tilde{0}},\ket{\tilde{1}}\}$ basis defined in equation \ref{eq:zero_one_tilde_bias} is that except for at $\theta=\pi/2$, where $\ket{\tilde{0}}=\ket{0}$ and $\ket{\tilde{1}}=\ket{1}$, the action of $Z$ is not simple anymore. We find that 
\begin{equation}
Z\rightarrow B_\mathbb{1}(\phi,\theta)\mathbb{1}+B_{\tilde{Z}}(\phi,\theta)\tilde{Z}+B_{\tilde{X}}(\phi,\theta)\tilde{X} \label{eq:Z_eff_bias}
\end{equation}
where 
\begin{align}
    B_\mathbb{1}(\phi,\theta)=\left(\cos^2(\phi)-\sin^2(\phi)\right)\left( \sin^2(\theta)-1\right) \\
    B_{\tilde{Z}}(\phi,\theta)= \left(\cos^2(\phi)-\sin^2(\phi)\right)^2\left( \sin^2(\theta)-1\right)+4\sin(\theta)\sin^2(\phi)\cos^2(\phi)\\
    B_{\tilde{X}}(\phi,\theta)=\left(\cos^2(\phi)-\sin^2(\phi)\right)\left( \sin^2(\theta)+1-2\sin(\theta)\right)\cos(\phi)\sin(\phi).
\end{align}

Since $B_\mathbb{1}(\phi,\pi/2)=B_{\tilde{X}}(\phi,\pi/2)=0$ and  $B_{\tilde{Z}}(\phi,\pi/2)=1$ independent of $\phi$, using fixed Ising interactions will always end with the system in an Ising Hamiltonian. Let us assume always-on interactions of the form
\begin{equation}
    H_{\mathrm{Ising}}=\sum_{j<k}J_{j,k}Z_jZ_k+Z_a\sum_j h_jZ_j,
\end{equation}
 We see from equation \ref{eq:Z_eff_bias} that a quadratic $Z_jZ_k$ term will result in nine terms all products of $\mathbb{1},\tilde{Z},\tilde{X}$. Of these terms the $\tilde{Z}_j\tilde{Z}_k$ terms are the desired interactions, while all terms involving $\tilde{X}$ will effectively act as additional driving terms and there will be an irrelevant global phase shift proportional to $B_\mathbb{1}^2(\theta)$. There will however also be diagonal terms which will distort the solution space, $\tilde{Z}_{j,k}$ terms which come with a prefactor of $B_\mathbb{1}(\theta)B_{\tilde{Z}}(\theta)$. Since these terms effectively change the diagonal part of the Hamiltonian, corresponding to the ``problem'' which is being solved, they may be desirable to remove. Fortunately these single $Z$ terms can be removed by adding $\theta$-dependant $Z$ terms specifically,
\begin{equation}
    H_{\mathrm{corr}}=-\sum_{j<k}J_{j,k}\frac{1}{2}\left(B_\mathbb{1}(\phi_j,\theta)Z_k+B_\mathbb{1}(\phi_k,\theta)Z_j\right),
\end{equation}
under the assumption that $\phi_a=\pi/4$ since there is no benefit  no overall effect from biasing this variable.

A final note for the biased driving definition is that a positive $\alpha$ value does not gaurantee that $\bigotimes_j\ket{u}_j$ will be the ground state, at $\theta=0$. At this point in the evolution the basis can be defined as $\{\ket{u},\ket{\zeta_-}\}$, but in general $\sandwich{\zeta_-}{Z}{\zeta_-}\neq 0$. However, in $\bigotimes_j\ket{u}_j$ will always be the ground state at $\theta=0$ as long as 
\begin{equation}
    \alpha>-\min_j\left[\sandwich{\zeta_-(\phi_j)}{Z}{\zeta_-(\phi_j)}\sum_{j,k}\frac{1}{2}J_{j,k}\sandwich{\zeta_-(\phi_k)}{Z}{\zeta_-(\phi_k)}\right].
\end{equation}
Alternatively, it is always possible to sweep both $\theta$ and $\phi$ during the protocol, starting with $\phi_j=\pi/4 \quad \forall j$.

\subsection{Qudit Variables \label{sub:qudits}}

A natural generalisation of the approach we have taken is to consider qudit variables. This can be achieved by coupling a larger system to a variable of the form given in equation \ref{eq:coupling_state_sup}.

\begin{equation}
    \ket{\xi(\theta)}=-\sin(\theta)\ket{u}+\cos(\theta)\ket{+}, \label{eq:coupling_state_sup}
\end{equation}

In this case, the state $\ket{+}\rightarrow \ket{\omega}$ will generalise to
\begin{equation}
    \ket{\omega}=\frac{1}{\sqrt{m}}\sum^{m-1}_{j=0} \ket{j}, \label{eq:qudit_plus}
\end{equation}
leading to,
\begin{equation}
    \ket{\xi(\theta)}=-\sin(\theta)\ket{u}+\cos(\theta)\ket{\omega}. \label{eq:coupling_state_qudit}
\end{equation}

In this case the orthogonal complement to $\ket{\xi(\theta)}$ will consist of a symmetric state and the space spanned by $m-1$ orthogonal state vectors. The symmetric state is identical to $\ket{\tilde{+}}$ in equation \ref{eq:allowed_span_sup}, 
\begin{equation}
    \{\ket{\tilde{+}}=\cos(\theta)\ket{u}+\sin(\theta)\ket{+},\ket{\tilde{-}}=\ket{-}\},\label{eq:allowed_span_sup}
\end{equation}
but using the new definition of $\ket{\omega}$ from equation \ref{eq:qudit_plus}
\begin{equation}
    \ket{\tilde{\omega}}=\cos(\theta)\ket{u}+\sin(\theta)\ket{\omega}.\label{eq:omega_tilde}
\end{equation}

To define an effective Hamiltonian, we define the closest allowed state within the subspace to state $\ket{j}$ we first need to define an anti-symmetrised version of $\ket{j}$,
\begin{align}
    \ket{\underline{j}}=
    \frac{1}{\mathcal{N}}\left(\ket{j}-\frac{1}{m-1}\sum_{k\neq j}\ket{k}\right)\\
    \mathcal{N}=\frac{1}{\sqrt{1+\frac{1}{m-1}}}
\end{align}
which has the key feature of being antisymmetric $\braket{\underline{j}}{\omega}=0$, but at the price of no longer being mutually orthogonal $\braket{\underline{j}}{\underline{k}}\neq 0$.
A useful identity that 
\begin{align}
    \ket{j}=\frac{1}{\mathcal{N}}\left(\ket{\omega}+\sqrt{\frac{1}{m}+\frac{1}{m(m-1)}}\ket{\underline{j}}\right),\\
    \mathcal{N}=\frac{1}{\sqrt{1+\frac{1}{m}+\frac{1}{m(m-1)}}}.
\end{align}
Equipped with this identity we define
\begin{align}
    \ket{\tilde{j}}=\frac{1}{\mathcal{N}}\left( \cos(\theta)\ket{u}+\sin(\theta)\ket{\omega}+\sqrt{\frac{1}{m}+\frac{1}{m(m-1)}}\ket{\underline{j}} \right) \label{eq:j_tilde}\\
    \mathcal{N}=\frac{1}{\sqrt{1+\frac{1}{m}+\frac{1}{m(m-1)}}}.
\end{align}
A first observation is that if we set $m=2$, equation \ref{eq:j_tilde} reduces to equation \ref{eq:zero_one_tilde_sup}.
\begin{align}
 \{\ket{\tilde{0}}=\frac{1}{\sqrt{2}}\left(\cos(\theta)\ket{u}+\sin(\theta)\ket{+}+\ket{-}\right),\nonumber \\
 \ket{\tilde{1}}=\frac{1}{\sqrt{2}}\left(\cos(\theta)\ket{u}+\sin(\theta)\ket{+}-\ket{-}\right)\}   \label{eq:zero_one_tilde_sup}
\end{align}
Returning to the general case, we note that if we apply an offset energy to the undefined state of the form \ref{eq:H_off_sup},
\begin{equation}
    H_{\mathrm{off}}=-\alpha \sum_j\ketbra{u_j}{u_j}.,\label{eq:H_off_sup}
\end{equation}
this effectively applies off-diagonal Hamiltonian terms linking all qudit states symmetrically, introducing Hamiltonian terms of the form
\begin{equation}
    H_\mathrm{qudit\,drive}=-\alpha \cos^2(\theta) \ketbra{\tilde{\omega}}{\tilde{\omega}}=-\frac{\alpha \cos^2(\theta)}{m+1+\frac{1}{(m-1)}}\sum_{j,k}\ketbra{\tilde{j}}{\tilde{k}} \label{eq:qudit_drive}
\end{equation}

\subsection{Higher Order Driving}

The next generalisation we will consider is a generalisation to higher order driving, in other words flipping multiple qubits or qudits simultaneously. The simplest version of such drivers follows immediately from section \ref{sub:qudits}; if multiple qubits are treated like a single qudit and coupled to the same undefined state, then this will create a driving term which couples between all possible states of the qubits. 

As a concrete example, consider
\begin{equation}
    \ket{\xi(\theta)}=-\sin(\theta)\ket{u}+\cos(\theta)\ket{{+}{+}}
\end{equation}
where $\ket{{+}{+}}$ indicates the $\ket{+}$ state over two different qubits. It follows directly from equation \ref{eq:qudit_drive} (plugging in $m=4$ and the four states within the qudit space), that the effective driving will be
\begin{align}
    H_\mathrm{pair\,drive}=-\sqrt{\frac{3}{64}}\alpha \cos(\theta) \nonumber \\
    \left(\ket{\tilde{0}\tilde{0}}+\ket{\tilde{0}\tilde{1}}+\ket{\tilde{1}\tilde{0}}+\ket{\tilde{1}\tilde{1}}\right) \left(\bra{\tilde{0}\tilde{0}}+\bra{\tilde{0}\tilde{1}}+\bra{\tilde{1}\tilde{0}}+\bra{\tilde{1}\tilde{1}}\right).\label{eq:pair_drive}
\end{align}
More commonly, higher order driving is considered in terms of Pauli $X$ strings, products of multiple $X_j$ terms. Such terms can be produced by coupling all computational basis states with an odd (or equivalently even) number of $\ket{+}$ terms to an undefined state which applies a phase. For example (up to an irrelevant identity factor) an $XX$ coupling could be achieved by forbidding 
\begin{align}
    \ket{\xi_{{+}{-}}(\theta)}=-\sin(\theta)\ket{u_{{+}{-}}}+\cos(\theta)\ket{{+}{-}} \\
    \ket{\xi_{{-}{+}}(\theta)}=-\sin(\theta)\ket{u_{{-}{+}}}+\cos(\theta)\ket{{-}{+}}
\end{align}
and applying appropriate phases to $\ket{u_{{+}{-}}}$ and $\ket{u_{{-}{+}}}$ as these are a pair of eigenstates which get a different phase from the symmetric pair. A disadvantage of this coupling method is that the number of states with an odd number of $\ket{+}$ states grows exponentially with the length of the string, in other words for a Pauli string of length $m$, $2^{m-1}$ such pairs of forbidden and undefined states would be required to implement it. 

Fortunately, there is a more efficient construction which takes advantage of the natural tensor product structure within Pauli strings. If we consider having instead of $2^{m-1}$ forbidden states which create a coupling to $2^{m-1}$ undefined states, we consider using these forbidden states to couple to a space of $m$ qubits. Since these have added $2^m$ states to the Hilbert space, to conserve the size of the space (and create an valid effective Hamiltonian) we will need to forbid $2^m$ states. Fortunately in this case we can also take advantage of the tensor product structure of the interaction, and therefore achieve this using $2 m$ forbidden states. To do this, for each pair of computational qudit and additional qubit used to build the undefined space we forbid\footnote{Note that there is a direct sum rather than tensor product structure between the forbidden states and the logical qubit states, not a tensor product structure, but there is a tensor product structure within each space} 
\begin{align}
    \ket{\xi_{0{+},j}(\theta)}=-\sin(\theta)\ket{0_{u,j}}+\cos(\theta)\ket{+_j}, \\
    \ket{\xi_{1{-},j}(\theta)}=-\sin(\theta)\ket{1_{u,j}}+\cos(\theta)\ket{-_j},
\end{align}
where the $u$ subscript indicates a term acting on the additional qubits used to build the space of undefined states. Forbidding these two states means that the effective two variable basis becomes 
\begin{align}
\ket{\tilde{+}}=\cos(\theta)\ket{0_{u,j}}+\sin(\theta)\ket{+_j}, \\
    \ket{\tilde{-}}=\cos(\theta)\ket{1_{u,j}}+\sin(\theta)\ket{-_j},
\end{align}
An important aspect of this structure is that a $\ket{0_{u,j}}$ is occupied iff $\ket{+_j}$ is occupied and vice-versa. Since a Pauli string of $X$ terms can be implemented by selectively applying differing phases to states with odd or even numbers of $\ket{+}$ or $\ket{-}$ on each qubit, this implies that applying a string of Pauli $Z_{u,j}$ operators will effectively implement an $X$ Pauli string. While such an implementation still requires a long Pauli string to be implemented, it is applied within the computational basis, which is considered easier to implement in many settings, for example through specially built gadgets \cite{chancellor16a,Leib16a,chancellor17a,Dodds2019permutation}.

For Pauli string type drivers to be useful, it should be possible to overlap them. Fortunately in the structure discussed here, there is a one-to-one mapping between $Z$ Pauli strings on the $u$ subspace and $X$ Pauli strings on the subspace of computational variables for example $-Z_{u,1}-Z_{u,2}+Z_{u,1}Z_{u,2}$ will implement a two body $XX$ coupling along with a transverse field on qubits $1$ and $2$. As this example demonstrates, a frustrated combination of $Z$ terms will implement a non-stoquastic driver, which have been an active area of research within quantum annealing \cite{Seki2012nonStoq,Seoane2012nonStoq,Seki2015nonStoq,albash16a,Ozfidian2020nonStoq,Vinci2017nonStoq}.

\section{Three satisfiability problem used within this paper \label{app:3SAT}}

This appendix provides the specific randomly generated three satisfiability problem which we have used in this study for reproducability purposes. Equation  \ref{eq:three_sat_CNF} gives the clauses in a standard conjunctive normal form.

\begin{align} 
 	\left( \neg x_{2}  \vee \neg x_{1}  \vee \neg x_{3}  \right) \wedge \left( \neg x_{1}  \vee \neg x_{3}  \vee \neg x_{2}  \right) \wedge \left( \neg x_{0}  \vee x_{1}  \vee \neg x_{2}  \right) \wedge  \nonumber \\ 
 	\left( \neg x_{2}  \vee x_{3}  \vee \neg x_{4}  \right) \wedge \left( \neg x_{4}  \vee \neg x_{0}  \vee \neg x_{3}  \right) \wedge \left( \neg x_{1}  \vee \neg x_{0}  \vee \neg x_{4}  \right) \wedge  \nonumber \\ 
 	\left( \neg x_{3}  \vee \neg x_{0}  \vee \neg x_{1}  \right) \wedge \left( \neg x_{1}  \vee x_{4}  \vee \neg x_{0}  \right) \wedge \left( \neg x_{2}  \vee \neg x_{1}  \vee x_{0}  \right) \wedge  \nonumber \\ 
 	\left( \neg x_{0}  \vee \neg x_{4}  \vee \neg x_{3}  \right) \wedge \left( \neg x_{4}  \vee \neg x_{2}  \vee x_{3}  \right) \wedge \left( \neg x_{3}  \vee \neg x_{1}  \vee \neg x_{0}  \right) \wedge  \nonumber \\ 
 	\left( \neg x_{1}  \vee \neg x_{2}  \vee x_{0}  \right) \wedge \left( \neg x_{3}  \vee \neg x_{2}  \vee x_{0}  \right) \wedge \left( \neg x_{4}  \vee x_{1}  \vee \neg x_{3}  \right) \wedge  \nonumber \\ 
 	\left( \neg x_{1}  \vee x_{0}  \vee \neg x_{4}  \right) \wedge \left( \neg x_{3}  \vee x_{2}  \vee x_{4}  \right) \wedge \left( \neg x_{3}  \vee x_{2}  \vee \neg x_{4}  \right) \wedge  \nonumber \\ 
 	\left( \neg x_{1}  \vee \neg x_{3}  \vee x_{4}  \right) \wedge \left( \neg x_{2}  \vee \neg x_{4}  \vee \neg x_{1}  \right) \wedge \left( \neg x_{3}  \vee \neg x_{1}  \vee x_{2}  \right) \wedge  \nonumber \\ 
 	\left( \neg x_{2}  \vee x_{1}  \vee \neg x_{4}  \right) \wedge \left( \neg x_{4}  \vee \neg x_{0}  \vee \neg x_{1}  \right) \wedge \left( \neg x_{3}  \vee x_{4}  \vee x_{0}  \right) \wedge  \nonumber \\ 
 	\left( \neg x_{2}  \vee x_{4}  \vee x_{0}  \right) \wedge \left( \neg x_{2}  \vee \neg x_{4}  \vee x_{3}  \right) \wedge \left( \neg x_{1}  \vee x_{4}  \vee \neg x_{2}  \right) \wedge  \nonumber \\ 
 	\left( \neg x_{4}  \vee x_{3}  \vee \neg x_{1}  \right) \wedge \left( \neg x_{2}  \vee x_{0}  \vee \neg x_{4}  \right) \wedge \left( \neg x_{3}  \vee \neg x_{0}  \vee x_{2}  \right) \wedge  \nonumber \\ 
 	\left( \neg x_{4}  \vee \neg x_{3}  \vee x_{2}  \right) \wedge \left( \neg x_{0}  \vee \neg x_{2}  \vee \neg x_{4}  \right) \wedge \left( \neg x_{0}  \vee \neg x_{4}  \vee x_{2}  \right) \wedge  \nonumber \\ 
 	\left( \neg x_{1}  \vee x_{4}  \vee x_{0}  \right) \wedge \left( \neg x_{4}  \vee \neg x_{2}  \vee x_{1}  \right) \wedge \left( \neg x_{2}  \vee \neg x_{3}  \vee \neg x_{4}  \right) \wedge  \nonumber \\ 
 	\left( \neg x_{0}  \vee \neg x_{4}  \vee \neg x_{1}  \right) \wedge \left( \neg x_{2}  \vee \neg x_{0}  \vee x_{4}  \right) \wedge \left( \neg x_{3}  \vee x_{2}  \vee \neg x_{1}  \right) \wedge  \nonumber \\ 
 	\left( \neg x_{0}  \vee \neg x_{2}  \vee x_{1}  \right) \wedge \left( \neg x_{0}  \vee \neg x_{3}  \vee \neg x_{1}  \right) \wedge \left( \neg x_{1}  \vee \neg x_{4}  \vee \neg x_{2}  \right) \wedge  \nonumber \\ 
 	\left( \neg x_{0}  \vee x_{2}  \vee x_{3}  \right) \wedge \left( \neg x_{2}  \vee \neg x_{1}  \vee x_{4}  \right) \wedge \left( \neg x_{4}  \vee x_{1}  \vee x_{0}  \right)
 \label{eq:three_sat_CNF}
 \end{align} 

For the reader's convenience we also provide this data in a more machine-readable format, equation \ref{eq:three_sat_var_LoL} gives the variable numbers for each clause as a python list-of-lists., while equation \ref{eq:three_sat_neg_LoL} lists whether or not each variable is negated, with $1$ corresponding to negation and $0$ corresponding to no negation. 
 
\begin{align} 
 	[ [ 2 , 1 , 3]  , [ 1 , 3 , 2]  , [ 0 , 1 , 2]  , [ 2 , 3 , 4]  , [ 4 , 0 , 3]  , [ 1 , 0 , 4]  , [ 3 , 0 , 1]  , [ 1 , 4 , 0]  , [ 2 , 1 , 0]  ,  \nonumber \\ 
 	[ 0 , 4 , 3]  , [ 4 , 2 , 3]  , [ 3 , 1 , 0]  , [ 1 , 2 , 0]  , [ 3 , 2 , 0]  , [ 4 , 1 , 3]  , [ 1 , 0 , 4]  , [ 3 , 2 , 4]  , [ 3 , 2 , 4]  ,  \nonumber \\ 
 	[ 1 , 3 , 4]  , [ 2 , 4 , 1]  , [ 3 , 1 , 2]  , [ 2 , 1 , 4]  , [ 4 , 0 , 1]  , [ 3 , 4 , 0]  , [ 2 , 4 , 0]  , [ 2 , 4 , 3]  , [ 1 , 4 , 2]  ,  \nonumber \\ 
 	[ 4 , 3 , 1]  , [ 2 , 0 , 4]  , [ 3 , 0 , 2]  , [ 4 , 3 , 2]  , [ 0 , 2 , 4]  , [ 0 , 4 , 2]  , [ 1 , 4 , 0]  , [ 4 , 2 , 1]  , [ 2 , 3 , 4]  ,  \nonumber \\ 
 	[ 0 , 4 , 1]  , [ 2 , 0 , 4]  , [ 3 , 2 , 1]  , [ 0 , 2 , 1]  , [ 0 , 3 , 1]  , [ 1 , 4 , 2]  , [ 0 , 2 , 3]  , [ 2 , 1 , 4]  , [ 4 , 1 , 0]  ]
 \label{eq:three_sat_var_LoL}
 \end{align} 

\begin{align} 
 	[ [ 1 , 1 , 1]  , [ 1 , 1 , 1]  , [ 1 , 0 , 1]  , [ 1 , 0 , 1]  , [ 1 , 1 , 1]  , [ 1 , 1 , 1]  , [ 1 , 1 , 1]  , [ 1 , 0 , 1]  , [ 1 , 1 , 0]  ,  \nonumber \\ 
 	[ 1 , 1 , 1]  , [ 1 , 1 , 0]  , [ 1 , 1 , 1]  , [ 1 , 1 , 0]  , [ 1 , 1 , 0]  , [ 1 , 0 , 1]  , [ 1 , 0 , 1]  , [ 1 , 0 , 0]  , [ 1 , 0 , 1]  ,  \nonumber \\ 
 	[ 1 , 1 , 0]  , [ 1 , 1 , 1]  , [ 1 , 1 , 0]  , [ 1 , 0 , 1]  , [ 1 , 1 , 1]  , [ 1 , 0 , 0]  , [ 1 , 0 , 0]  , [ 1 , 1 , 0]  , [ 1 , 0 , 1]  ,  \nonumber \\ 
 	[ 1 , 0 , 1]  , [ 1 , 0 , 1]  , [ 1 , 1 , 0]  , [ 1 , 1 , 0]  , [ 1 , 1 , 1]  , [ 1 , 1 , 0]  , [ 1 , 0 , 0]  , [ 1 , 1 , 0]  , [ 1 , 1 , 1]  ,  \nonumber \\ 
 	[ 1 , 1 , 1]  , [ 1 , 1 , 0]  , [ 1 , 0 , 1]  , [ 1 , 1 , 0]  , [ 1 , 1 , 1]  , [ 1 , 1 , 1]  , [ 1 , 0 , 0]  , [ 1 , 1 , 0]  , [ 1 , 0 , 0]  ]
 \label{eq:three_sat_neg_LoL}
 \end{align} 

\section{Satisfiability\label{sec:satisfiabilty}}

We now consider what happens if in addition to projecting out the $\ket{\xi(\theta)}$ state using a Zeno effect, we also prevent the system from entering classical states which do not satisfy a given constraint. One canonical problem in this direction which is often used in proofs of NP-completeness is the problem of three satisfiability, the question of whether an arrangement of variables can be found which satisfy the following conjunctive normal form (CNF) statement,
\begin{equation}
    \bigwedge_{j,k,l \in \mathrm{clauses}} (x_j\vee x_k \vee x_l),
\end{equation}
where indexes are allowed to be positive or negative with the understanding that $x_{-j}=\neg x_j$. This formula uses the standard notation where $\vee$ indicates disjunction (OR), $\wedge$ indicates conjunction (AND) boolean variable, $x\in \{\mathrm{TRUE} \equiv 1,\mathrm{FALSE} \equiv 0\}$ . The arguments we give here will work for any constraint satisfaction problem, but this provides a concrete example.

Satisfiability solvers are generally divided into two types \cite{biere2009SAT,Gong2017SAT}; incomplete solvers, which find a satisfying arrangement if one exists, and complete solvers, which are also capable of certifying that no satisfying assignment exists. When applied to satisfiability problems, quantum annealing provides an incomplete solver, if no satisfying assignment exists even a perfectly adiabatic protocol would find a state which satisfies the maximum number of clauses but would give no indication of whether or not a higher quality solution exists. We focus mostly on constructing an incomplete solver in this work, but do discuss that dissipative protocols potentially provide an avenue toward complete solvers, which does not exist for annealing protocols.

A clause in a statisfiability problem effectively forbids a subset of variables from taking a given set of values, for instance $x_1\vee x_2 \vee \neg x_3$, is exactly equivalent to forbidding cases where simultaneously $x_1=0$, $x_2=0$, and $x_3=1$. Such a statement can therefore naturally be represented by coupling strongly to $\ket{0}_1\ket{0}_2\ket{1}_3$ and therefore preventing a system from entering such a state regardless of the states of other qubits.

A Zeno interaction which forces a system to remain within the satisfiable states of a three (k) statisfiability problem can therefore naturally be represented by a sum of three (k) body coupling terms which couple forbidden states strongly to an environment. If we wish to combine this with the formalism of section 2 of the main text, we need to additionally define the action of a clause on the undefined $\ket{u}$ state. A natural choice is to define that a clause containing at least one variable in the $\ket{u}$ state is satisfied iff the $u$ variable(s) could be assigned values which would satisfy the clause. For clauses written as a CNF, this is equivalent to stating that a clause containing at least one $\ket{u}$ variable is always satisfied, but this will not always be the case for other kinds of constraints.

With this behaviour defined we note that iff a satisfying arrangement with bit values $s_i$ exists the state
\begin{align}
    \ket{\phi(\theta)}=\frac{1}{\left(1+\sin^2(\theta)\left(\sqrt{2}-1\right)^2\right)^{\frac{n}{2}}}\bigotimes_{i=1}^{n} \left(\cos(\theta) \ket{u}+\sqrt{2}\sin(\theta) \ket{s_i}\right) \label{eq:sat_state_s}\\
    =\frac{1}{\left(1+\sin^2(\theta)\right)^{\frac{n}{2}}}\bigotimes_{i=1}^{n}  \left(\cos(\theta) \ket{u}+\sin(\theta)\left(\ket{+}+(-1)^{s_i}\ket{-}\right) \right)\label{eq:sat_state_pm} \\
    =\frac{1}{\left(1+\sin^2(\theta)\right)^{\frac{n}{2}}}\bigotimes_{i=1}^{n} \left(\ket{\tilde{+}}+(-1)^{s_i}\sin(\theta)\ket{\tilde{-}}\right) \\
    =\frac{1}{\left(2+2\sin^2(\theta)\right)^{\frac{n}{2}}}\bigotimes_{i=1}^{n}\left((1+(-1)^{s_i}\sin(\theta))\ket{\tilde{0}}+(1-(-1)^{s_i}\sin(\theta))\ket{\tilde{1}}\right)
\end{align}
will remain within the subspace of allowed states. This can be seen because on one hand, by definition all clauses are satisfied. On the other hand for each qubit the variable lies in the space defined by equation \ref{eq:allowed_span_sup}.Where no satisfying assignment exists, it is not possible to construct such a state for $\theta \neq 0$. If multiple satisfying assignments exist than any superposition of states of the form given in equation \ref{eq:sat_state_s} will be eigenstates with a zero eigenvalue. Note that this assumes that slow enough evolution is possible to effectively implement a Zeno effect, if this is not the case, the solver will fail, and will either be found in a decayed state, or an excited state corresponding to an assignment which fails to satisfy the clauses, depending on the kind of Zeno effect being implemented.

\begin{figure}
    \centering
    \includegraphics[width=5cm]{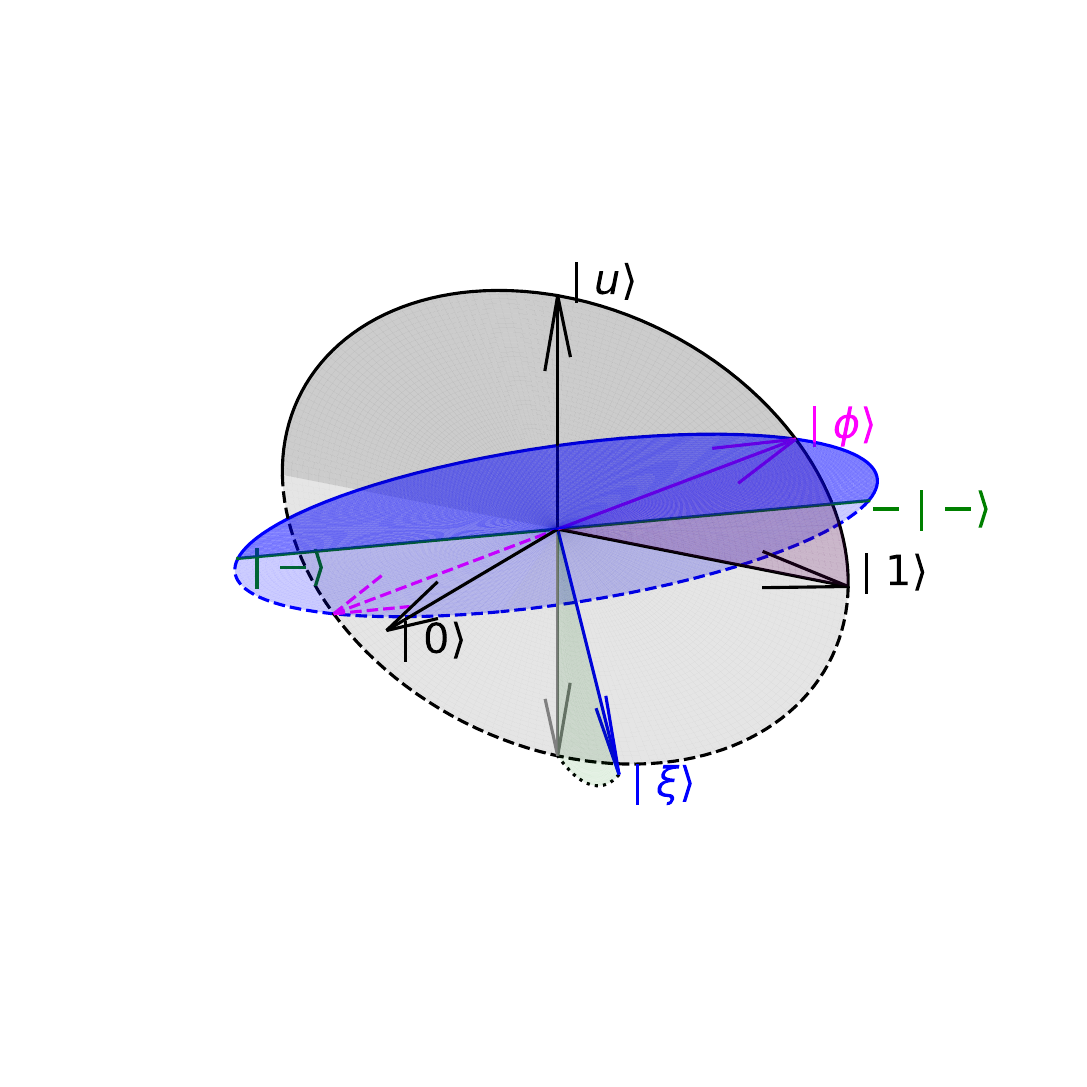}\includegraphics[width=5cm]{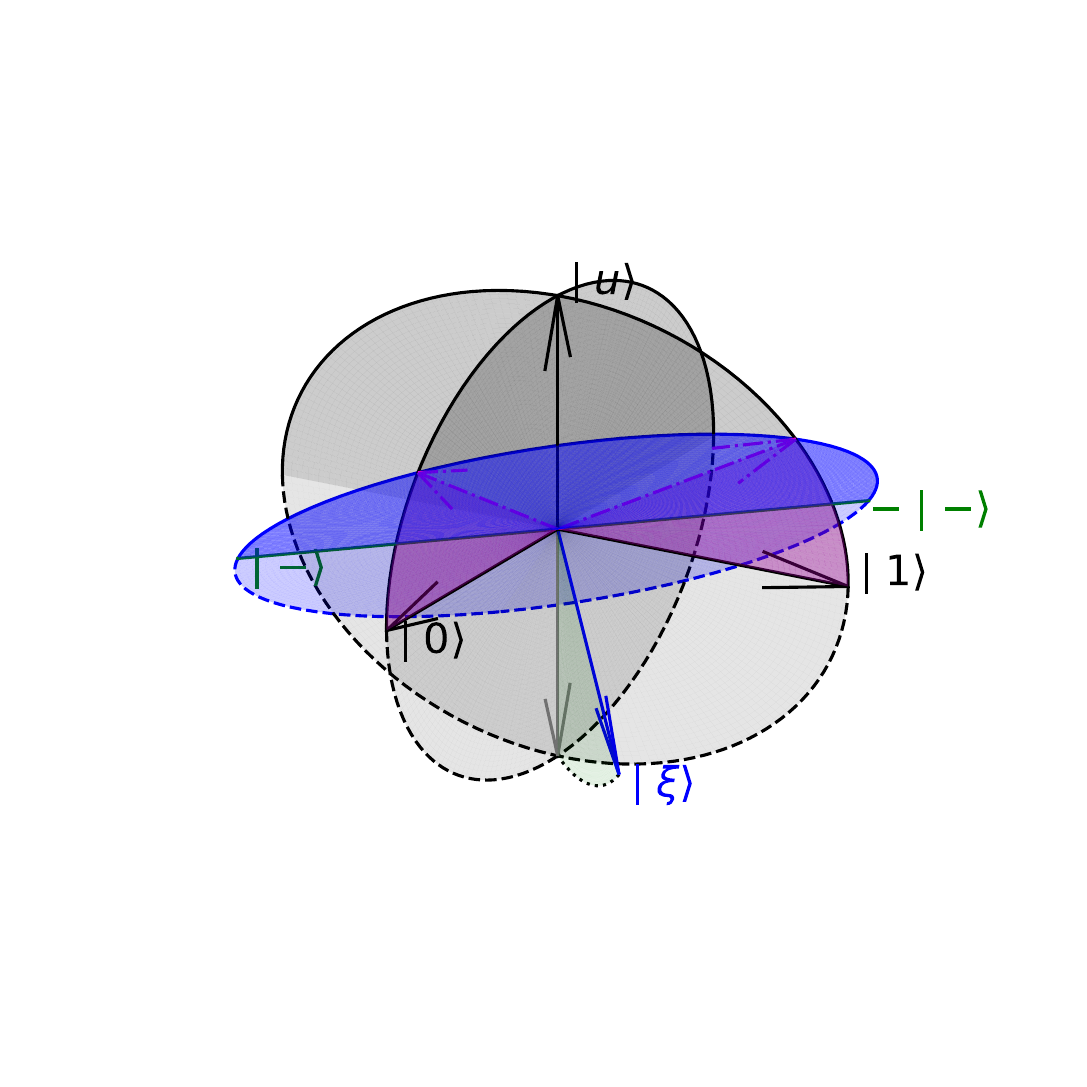}
    \caption{Left: Visualisation of a $\ket{\phi}$ state which is constrained to be orthogonal to both $\ket{\xi}$ (visualised as blue disk) and $\ket{0}$ (visualised as grey disk) for $\theta=\pi/4$. Redundant parts of the space which differ only by a global phase are shown in a lighter colour. A second intersection vector corresponding to $-\ket{\phi}$ (indistinguishable under the laws of quantum mechanics) is shown using dashed lines. The dotted line and green shading are a guide to the eye. Right: visualisation that it is not possible to find a vector which is simultaneously orthogonal to $\ket{0}$, $\ket{1}$ (both spaces represented by grey disks) and $\ket{\xi}$.}
    \label{fig:forbid_0}
\end{figure}

As was done in figure 1 of the main text, we can visualise the additional effect of a combined satisfiability constraint along with forbidding a state $\ket{\xi}$. The result of forbidding $\ket{0}$ is shown in figure \ref{fig:forbid_0} where only a single state $\ket{\phi}$ is allowed, which lies at the intersection of the spaces orthogonal to $\ket{0}$ and $\ket{\xi}$. It is further apparent that it is not possible to obtain a state which is orthogonal to all three vectors.

Assuming a Zeno effect can be maintained while $\theta$ is changed continuously, the system will end in $\ket{\psi(\theta=\pi/2)}=\ket{s}$ (or a superposition of such states if multiple satisfying assignments exist). Assuming it runs perfectly this method is guaranteed to find a satisfying assignment in a similar nature to the guarantee of final optimality which adiabatic protocols have. If no satisfying assignment exists, then what happens will depend on the nature of the Zeno effect being used to implement the protocol. 

We consider the three families discussed in \cite{Berwald2024Zeno}. Firstly for a Zeno effect defined by phase flipping, adiabatic evolution or multi-stage quantum walks. Given restrictions of the space, the system would have to end in a state which is not a satisfying assignment and/or contains $\ket{u}$ somewhere. Similar arguments hold for effects driven by measurement or dephasing. Since it is easy to check if a state represents a satisfying assignment, these outcomes can easily be checked and discarded, and if they continue for very strong coupling or long runtimes, they suggest that a satisfying assignment does not exist. In the case of a Zeno effect based on destruction/dissipation, then the system will end in a ``destroyed'' state. The implications of such protocols will be discussed later, as they allow a method to check if the clauses are satisfiable and finding a satisfiable assignment without ever completing a sweep all the way to $\theta=\pi/2$. Examples of other constraints, and how they can be combined with interactions can be found in section \ref{sec:const_int_combine}.

\section{Iterative and hybrid satisfiability solvers \label{app:iterative_sat}}

It is worth considering what happens when a problem is satisfiable and a Zeno sweep has been used in such a way which we arrive at a value $0<\theta < \pi/2$. In this case each variable will exist in a superposition between $\ket{u}$ and a bit value which is part of a satisfying assignment. Assuming that the sweep has been performed slowly enough that the probability for the state to be lost can be neglected, then based on equation \ref{eq:sat_state_s} each variable will independently have a probability of being measured in the $\ket{u}$ state of 
\begin{equation}
    p_{\ket{u}}=\frac{\cos^2(\theta)}{1+\sin^2(\theta)(\sqrt{2}-1)^2}.
\end{equation}
The remaining probability $(1-p_{\ket{u}})$ is the probability of obtaining one bit of the satisfying assignment. The probability of obtaining every bit of the solution simultaneously is exponentially small in the number of variables $(1-p_{\ket{u}})^n$, but likewise the probability of obtaining no useful bits of information is $(p_{\ket{u}})^n$, so also exponentially rare. The most likely scenario is that a measurement gives a partial assignment, some $0$ and $1$ values which must belong to a satisfying assignment, while others are measured in an undefined state. Such partial assignments, or a collection of such partial assignments could be used within a classical algorithm, but multiple calls to the solver could be used to iteratively find a full satisfying assignment. 

It is worth emphasizing that this requires a strong assumption that slow enough evolution to avoid decay with a reasonable probability is possible. If decoherence or another technical issue prevented a sufficiently slow sweep from being performed, than the result would almost always be a decayed state, which does not give any bits of a satisfiable clause. For this reason the solver discussed here is unlikely to be technically feasible, but is interesting from a theoretical perspective. 

If we first perform a Zeno sweep to $0<\theta < \pi/2$ and then measure, then assuming large $n$ on average we obtain approximately $n_\mathrm{assign}\approx (1-p_{\ket{u}}) n$ bits of the solution (with the remainder measured in the $\ket{u}$ state), these bits can be fixed giving a new, smaller satisfiability problem of size $n-n_\mathrm{assign}$, such sweeps can be performed iteratively until all variables are assigned a value. Assuming perfect Zeno operation, the number of iterations for the a solver described here to fully converge will only grow linearly\footnote{Note that this is number of iterations and not \textbf{total runtime} which will scale linearly, as the latter would imply that $P=NP$, which is strongly suspected to be untrue, it is likely that exponential scaling lies elsewhere, for example the time to reliably perform each Zeno sweep.} with $(1-p_{\ket{u}})$, so therefore would be feasible even if implemented with a relatively small sweep in $\theta$. This contrast traditional annealing methods where a full transformation from a driver Hamiltonian to a problem Hamiltonian is needed, and is likely to be easier to achieve with many technologies. At each stage the algorithm could also effectively act as a satisfiability detector, allowing it to detect if an error exists in the assignment found in a previous iteration and therefore allowing backtracking to a partial assignment which is satisfiable.

While we have discussed a dissipative implementation of this algorithm, nothing would prevent an adiabatic implementation of this algorithm, in which case the equivalent of dissipation would be to instead enter an excited state which does not correspond to a satisfying assignment. The only difference here is that there would be no way to know if an error in fixing the variables has led to a problem which no longer has a satisfying assignment.

\section{Combining Constraints and Interactions \label{sec:const_int_combine}}

A natural consideration is what happens when constraints similar to those discussed in section \ref{sec:satisfiabilty} are combined with interactions as described in section 3 of the main text. Mathematically, this system becomes more complicated because a tensor product of the states defined in equation \ref{eq:allowed_span_sup} or equivalently equation \ref{eq:zero_one_tilde_sup} will not form a valid basis as these will contain all computational basis states, including those which violate the constraints. On the other hand, for each set of $\ket{0}$ and $\ket{1}$ values which satisfies the constraints, a state of the form given in equation \ref{eq:sat_state_s} will exist. 

For $0\le \theta<\pi /2$ states of the form given in equation \ref{eq:sat_state_s} for different satisfying arrangements will be mutually non-orthogonal, but for $0< \theta \le \pi /2$ will be linearly independent, meaning that they can be used to construct an orthogonal vector space with a dimension equal to the number of satisfying bitstrings. At the two extremes they will be identical at $\theta=0$ and corespond to the usual bitstrings at $\theta=\pi/2$.

While in principle any constraint can be cast in CNF form, this is often not the natural way to consider many which arise in the real world. We also consider constraints on the number of variables taking the $1$ value within a subset of the variables. Since optimisation problems based on allocating limited resources are common in the real world, constraints of this form come about naturally, and problems of optimising  cost subject to a constraint, or series of constraints are very common. 

\subsection{Example: one-hot and domain-wall constraints \label{sub:oh_dw}}

We start with two simple examples, a one-hot constraint which requires that exactly one variable takes the $1$ value and all others must take the $0$ value, and a related encoding known as the domain-wall encoding. One reason that one-hot constraint is particularly interesting is that it is not possible to pass between valid states using single bit flips, while in domain-wall encoding they do. Very naive intuition would suggest that since the unconstrained Hamiltonian implements an effective transverse field Ising model, adding a one-hot constraint would effectively prevent any interesting dynamics and therefore the system would not be able to find solutions. In this section we demonstrate both analytically and numerically that this naive intuition is not correct and rich dynamics capable of finding optimal solutions are present. While a system subject to a global one-hot constraint (or domain-wall encoding) does not have the robust tensor product structure required for scalable quantum computing \cite{Blume-Kohout02a}, a combination of multiple such subsystems does. This implies that to be scalable (and therefore interesting from a computing perspective), a system should be comprised of multiple domain-wall or one-hot systems, and not simply a single one. A single one-hot or domain-wall system would have a state space which only grows linearly with the number of qubits and therefore could efficiently be sampled classically. 

Considering constraints of this type is also physically well motivated. A common topic in studies on the subject of Zeno dynamics are systems where it is forbidden to have a specific number of excitations\cite{Signoles2014confined,Bretheau2015confined}. An alternative, which is less symmetric but more efficient in terms of variable count is to implement a domain-wall encoding \cite{chancellor2019,chen21a,Berwald2023Domain}. The domain-wall encoding constraint, where there is exactly one domain-wall where variables change from $1$ to $0$, when listed in order, assuming a ``virtual'' variable which takes the $1$ value before the chain and one which takes the $0$ value afterward can efficiently be expressed as a conjunction of CNF clauses, each with two literals
\begin{equation}
    \bigwedge^{m-1}_{j=1}(x_j\vee \neg x_{j+1}). \label{eq:domain_wall_CNF}
\end{equation}
The effect of such clauses is to forbid an anti-domain-wall where $x_j=0$ and $x_{j+1}=1$. The only states which do not have any anti-domain-walls are valid domain-wall-encoding states where for some $k$ $x_{j\le k}=1$ while $x_{j>k}=0$.

It immediately follows from equation \ref{eq:sat_state_s} that a solution space of $m+1$ states can be formed from superpositions of undefined $\ket{u}$ states and the $m+1$ states which obey the domain-wall constraint. Furthermore, if phases are applied to $\ket{u}$ states in the form given in equation \ref{eq:H_off_sup}, then the phase will be effectively applied to all $\ket{+}$ states, since domain-wall states are connected by single bit flips, this will mediate direct transitions between computational basis states. Specifically, if we order the domain wall states by Hamming weight, these will induce terms of the form $\ketbra{j}{j+1}+\ketbra{j}{j-1}$. There will also be additional mixing from the fact that different $\ket{\phi_j(\theta\neq \pi/2)}$ \ref{eq:sat_state_s} will not be orthogonal even if the corresponding $\ket{s}$ vectors are, as we explore later in the setting of one-hot encoding. 

\begin{figure}
    \centering
    \begin{subfigure}[b]{0.475\textwidth}
            \centering
            \includegraphics[width=\textwidth]{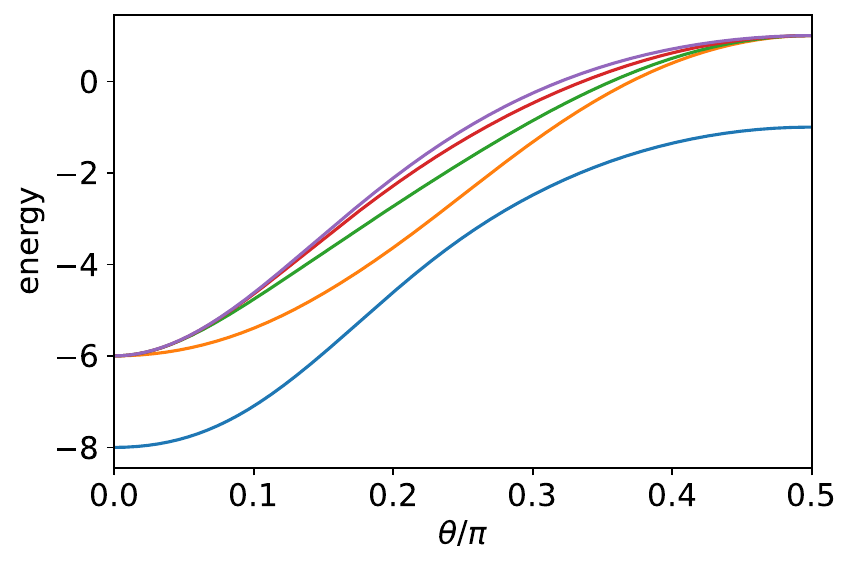}
            \caption[Spectrum of allowed states]%
            {Spectrum of allowed states}
            \label{fig:dw4_spectrum}
    \end{subfigure}
    \begin{subfigure}[b]{0.475\textwidth}
            \centering
            \includegraphics[width=\textwidth]{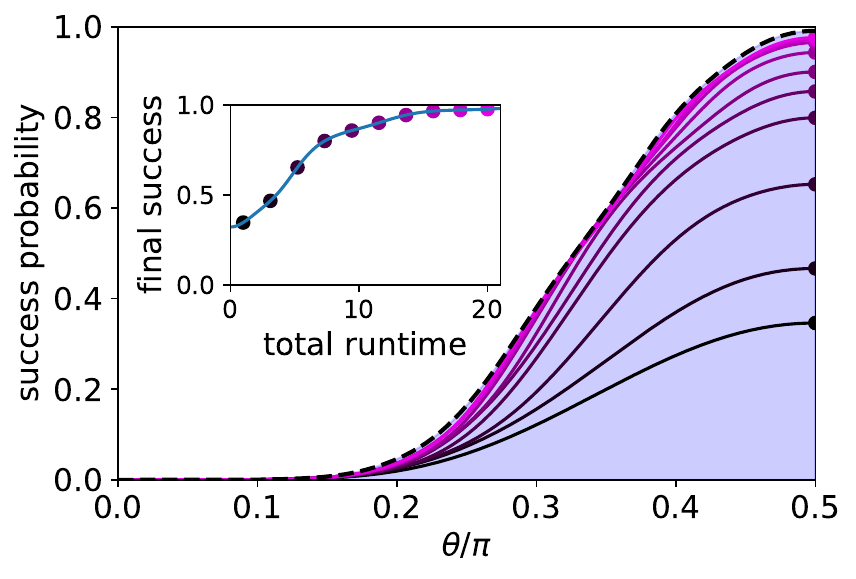}
            \caption[Success probabilities]%
            {Success probabilities}
            \label{fig:dw4_success}
     \end{subfigure}
    \caption{Spectrum (left) and success probability (right, defined as probability to be in the $\ket{1111}$ state). For a domain-wall system with four qubits and therefore five allowed domain-wall states. The shaded area bounded by a dashed line on the left plot represents the total probability to be in the manifold of valid domain-wall states, representing an upper bound on success probability during the sweep of $\theta$. The total runtime is colour coded in the lines and the final success probabilities are compared in the subfigure, which also gives total runtime. These plots use a bias term which is given in equation \ref{eq:H_off_sup} with $\alpha=2$ and the system starts with all variables in the $\ket{u}$ state. Perfect projection into the manifold of allowed states is assumed.}
    \label{fig:dw4_plots}
\end{figure}

Figure \ref{fig:dw4_plots} shows a numerically calculated example of annealing a single domain wall encoded variable with $5$ possible values where the $\ket{1111}$ is assigned an energy which is $2$ lower than all other configurations. These simulations start with $\alpha=2$ and assumes perfect projection into the allowed manifold of states, which are orthogonal to all $\ket{\xi(\theta)}$ states as well as configurations which violate the domain-wall conditions (contain any anti-domain-walls). A bias term of the form given in equation \ref{eq:H_off_sup} is applied with $\alpha=2$, and a linear sweep in $\theta$ used. As figure \ref{fig:dw4_spectrum} shows the system exhibits a spectrum which features an avoided level crossing, and as a result increasing the total runtime of a sweep leads to an approach to the adiabatic limit as depicted in figure \ref{fig:dw4_success}. For these numerical experiments we perform $1000$ Zeno projective measurements of whether or not the state is within the allowed state manifold, reducing the normalisation of the state vector for all states which are projected out.

One-hot constraints represent a case which is less intuitively clear a-priori because there are no single-bit-flip transitions between one-hot states. In the transverse field setting, this means that transitions must occur perturbatively, and for a perfect constraint where no states outside of the one-hot manifold are ever allowed, the dynamics completely stop.\footnote{The consequences of the difference between the perturbative and non-perturbative nature of one-hot and domain-wall encoding respectively can be seen experimentally in \cite{Berwald2023Domain}}. Naively one may assume this means that the transverse field intuition can be extrapolated and a model of the form considered in this paper. Based on this naive intuition, one-hot constraints would render the system unable to solve problems, we argue analytically and numerically however that this naive intuition is flawed and such constraints can be effectively used.

For notational simplicity we define the state $\ket{j}$ which corresponds to the $j$th variable taking a $1$ value and all others taking a value of $0$. We further define $\ket{\phi_j(\theta)}$ to be the non-orthogonal state vector corresponding to each of these assignments. From these non-orthogonal state vectors, we can define a basis of orthogonal state vectors using the permutation symmetry of the states\footnote{An astute reader will notice that this formula does not define a set of orthogonal vectors at exactly $\theta=0$ since $\ket{\phi_j(\theta=0)}$ are identical for all $j$, in practice we can avoid this special case by starting calculations at $\theta=\epsilon$.}
\begin{equation}
    \ket{\bar{\phi}_j(\theta)}=\frac{1}{\mathcal{N}}\left(\ket{\phi_j(\theta)}-a(\theta)\sum_{k\neq j}\ket{\phi_k(\theta)}\right),
\end{equation}
where $\mathcal{N}$ is a normalisation factor. The goal of constructing these states is to allow a space of orthogonal state vectors which tend toward $\ket{\phi_j(\theta=\pi/2)}$ in a symmetric way. The constant $a(\theta)$ can be derived from the orthogonality condition
\begin{align}
    \braket{\bar{\phi}_j(\theta)}{\bar{\phi}_{l\neq j}(\theta)}=0\\
    =\frac{1}{\mathcal{N}^2}\left(\braket{\phi_j(\theta)}{\phi_l(\theta)}-2a(\theta)+a^2(\theta) \sum_{k\neq j}\sum_{q\neq l}\braket{\phi_j(\theta)}{\phi_q(\theta)}\right)\\
    =\frac{1}{\mathcal{N}^2}\left( \braket{\phi_j(\theta)}{\phi_l(\theta)}-2a(\theta)+a^2(\theta)\left[n-2+(n-1)^2\braket{\phi_j(\theta)}{\phi_l(\theta)} \right] \right),
\end{align}
where the third line of the calculation follows by symmetry, and in the second line we have used the normalisation of $\ket{\phi_j(\theta)}$. Note that $n$ is the number of variables and therefore the number of states which satisfy the constraints. The value of $a(\theta)$ can be calculated using the quadratic formula 
\begin{equation}
    a(\theta)=\frac{1}{2 b(\theta)}\left(2 - \sqrt{4-4 b(\theta)\braket{\phi_j(\theta)}{\phi_{l\neq j}(\theta)}}\right)
\end{equation}
where 
\begin{equation}
    b(\theta)=n-2+(n-1)^2\braket{\phi_j(\theta)}{\phi_{l\neq j}(\theta)}.
\end{equation}
Note that we have taken the negative branch of the root in the quadratic formula because that is the branch which recovers $a(\pi/2)=0$ when $\braket{\phi_j(\pi/2)}{\phi_{l\neq j}(\pi/2)}=0$.

An important observation for understanding how this system will evolve as $\theta$ is changed is to observe that while $\braket{\bar{\phi}_j(\theta)}{\bar{\phi}_{l\neq j}(\theta)}=0$, we have
$\braket{\bar{\phi}_j(\theta)}{\bar{\phi}_{l\neq j}(\theta+\delta \theta)} \neq 0$. In other words the system will mix between the valid states as $\theta$ is swept. 

\begin{figure}
    \centering
    \begin{subfigure}[b]{0.475\textwidth}
            \centering
            \includegraphics[width=\textwidth]{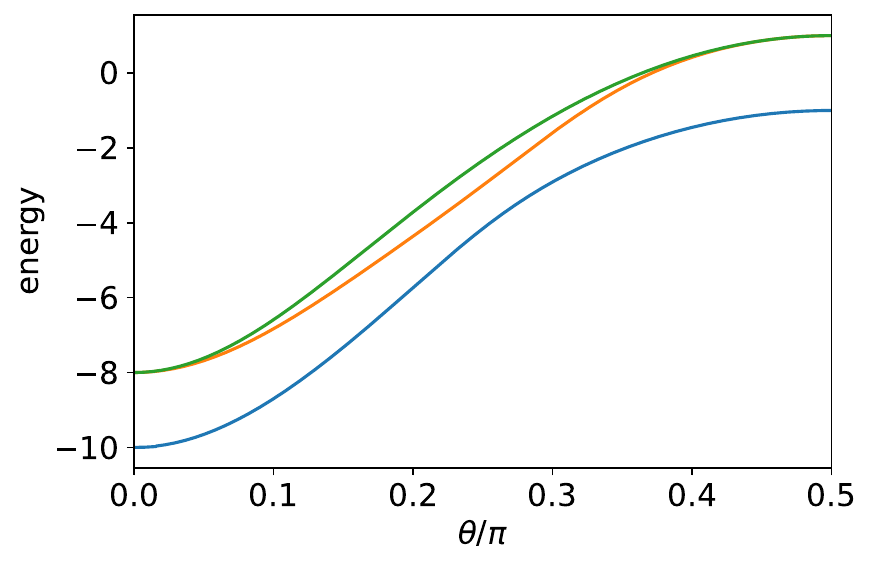}
            \caption[Spectrum of allowed states]%
            {Spectrum of allowed states}
            \label{fig:oh5_spectrum}
    \end{subfigure}
    \begin{subfigure}[b]{0.475\textwidth}
            \centering
            \includegraphics[width=\textwidth]{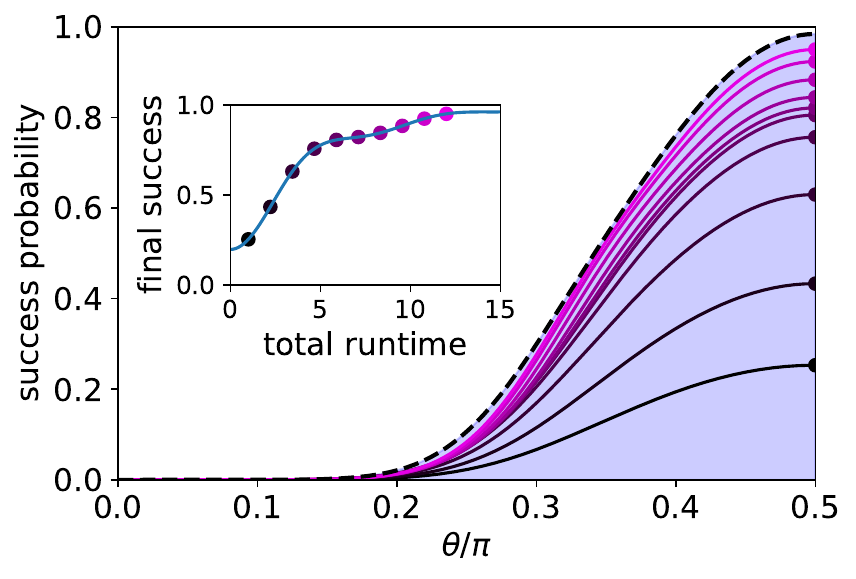}
            \caption[Success probabilities]%
            {Success probabilities}
            \label{fig:oh5_success}
     \end{subfigure}
    \caption{Spectrum (left) and success probability (right, defined as probability to be in the $\ket{10000}$ state). For a one-hot system with five qubits and therefore five allowed one-hot states. The shaded area bounded by a dashed line on the left plot represents the total probability to be in the manifold of valid one-hot states, representing an upper bound on success probability during the a sweep of $\theta$. The total runtime is colour coded in the lines and the final success probabilities are compared in the subfigure, which also gives total runtime. These plots use. A bias term given in equation \ref{eq:H_off_sup} is applied with $\alpha=2$ and the system starts with all variables in the $\ket{u}$ state. Perfect projection into the manifold of allowed states is assumed.}
    \label{fig:oh5_plots}
\end{figure}

The case where a one-hot constraint is combined with phases given by a non-trivial cost function would be complicated to approach analytically, but the the fact that an properly defined orthogonal space exists, combined with the fact that sweeping $\theta$ can mix within this subspace suggests that the adiabatic theorem of quantum mechanics should apply. The former condition guarantees that phases will lead to a well-defined Hamiltonian within the subspace, while the latter implies that the behaviour during the sweep will be defined by avoided, rather than true energy level crossings. The second fact can be confirmed by observing the dependence of off-diagonal elements in the $\ket{\bar{\phi}}$ basis on $\theta$, since 
\begin{equation}
    Z_l\ket{\bar{\phi}_l(\theta)}=\ket{\bar{\phi}_l(\theta)}-\frac{2\sqrt{2}\sin(\theta)}{\mathcal{N}\left(1+\sin^2(\theta)\left(\sqrt{2}-1\right)^2\right)^{\frac{n}{2}}}\ket{s_l},
\end{equation}
Where $\ket{s_l}$ is the vector corresponding to the classical solution, with $\ket{0}$ in all but the lth variable which takes a state of $\ket{1}$. We find that 
\begin{align}
    \sandwich{\bar{\phi}_j(\theta)}{Z_l}{\bar{\phi}_{l\neq j}(\theta)}=-\frac{2\sqrt{2}\sin(\theta)\braket{\bar{\phi}_j(\theta)}{s_l}}{\mathcal{N}\left(1+\sin^2(\theta)\left(\sqrt{2}-1\right)^2\right)^{\frac{n}{2}}}\\
    =\frac{4\sin^2(\theta)a(\theta)}{\mathcal{N}^2\left(1+\sin^2(\theta)\left(\sqrt{2}-1\right)^2\right)^{n}},
\end{align}
the presence of non-zero off-diagonal elements confirms that adiabatic state transfer will be facilitated. We confirm this behaviour numerically in figure \ref{fig:oh5_plots}, which plots the same quantities as \ref{fig:dw4_plots} but for the equivalent one-hot encoding. It demonstrates the theoretical prediction both that the spectrum is defined by an avoided rather than true crossing (figure \ref{fig:oh5_spectrum}) and that an adiabatic limit can be approached within the subspace (figure \ref{fig:oh5_success}). For these numerical experiments we perform $1000$ Zeno projective measurements of whether or not the state is within the allowed state manifold, reducing the normalisation of the state vector for all states which are projected out.

\subsection{Example: Resource Constraints}

A common constraint in the real world is a limitation on the number of some scarce resource \cite{Bradley1977AppliedMP}, so constraints of the form
\begin{equation*}
    \sum_jx_j \le C
\end{equation*}
are common. We refer to such constraints as resource constraints since they could for example come about where a limited number of some resource are allocated, similar constraints could also come about however from capacities of a system, as in knapsack type problems, although the interesting cases of these problems usually involve a weighted sum. Recall that such constraints also occur quite naturally in physical systems, and are often considered in examples of Zeno dynamics \cite{Signoles2014confined,Bretheau2015confined}. The combination of presence in computationally and commercially interesting problems and feasibility of physical implementation make constraints of this form a very natural choice to investigate further.

\begin{figure}
    \centering
    \begin{subfigure}[b]{0.475\textwidth}
            \centering
            \includegraphics[width=\textwidth]{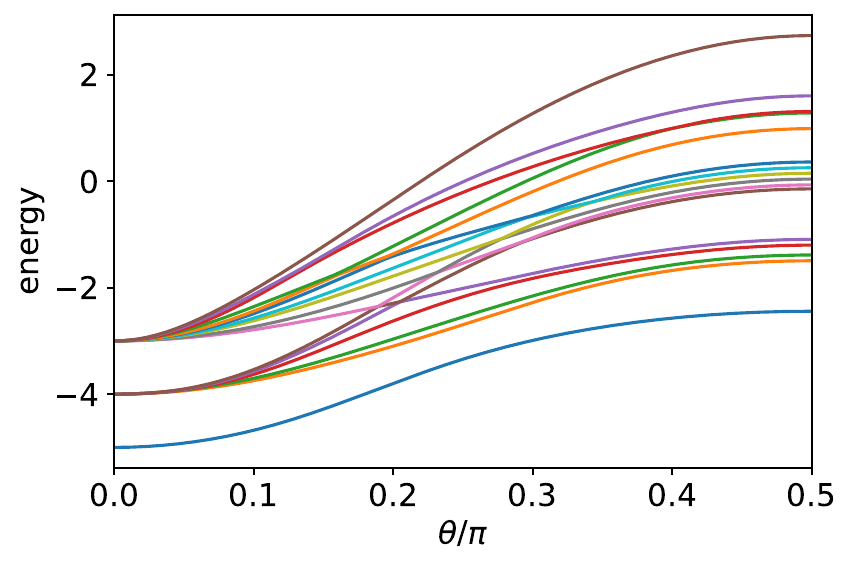}
            \caption[Spectrum of allowed states]%
            {Spectrum of allowed states}
            \label{fig:g2_spectrum}
    \end{subfigure}
    \begin{subfigure}[b]{0.475\textwidth}
            \centering
            \includegraphics[width=\textwidth]{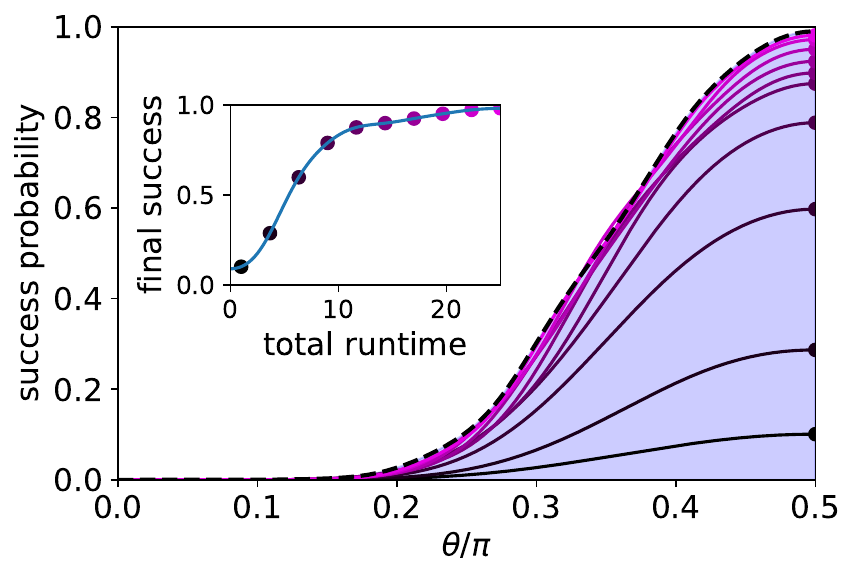}
            \caption[Success probabilities]%
            {Success probabilities}
            \label{fig:g2_success}
     \end{subfigure}
    \caption{Spectrum (left) and success probability (right). For a system with a random field Hamiltonian defined in equation \ref{eq:rand_h_sup} and a constraint that two or fewer of the qubits are allowed to take a zero value. The shaded area bounded by a dashed line on the right plot represents the total probability to be in the manifold of valid one-hot states, representing an upper bound on success probability during the sweep of $\theta$. The total runtime is colour coded in the lines and the final success probabilities are compared in the subfigure, which also gives total runtime. These plots use the bias term given in equation \ref{eq:H_off_sup} with $\alpha=1$ and the system starts with all variables in the $\ket{u}$ state. Perfect projection into the manifold of allowed states is assumed.}
    \label{fig:g2_plots}
\end{figure}

For these problems, the solution space lacks the extreme symmetry of one-hot-constrained systems, so they are less approachable analytically; however, we can gain valuable insights studying systems constrained in this way numerically. Figure \ref{fig:g2_plots} shows the results of such a run using random fields
\begin{equation}
\vec{h}=\left[ -0.67513783, -0.62099006, -0.14675767,  0.72688415,  0.56602992\right], \label{eq:rand_h_sup}
\end{equation}
and a constraint on the number of ones. As we see even for this more complicated constraint, the success probability quickly approaches unity.

\ifx \justbeingincluded \undefined

\bibliographystyle{unsrt}
\bibliography{references}  

\end{document}

\fi

\bibliography{references}  

\end{document}